\documentclass[%
twocolumn,
superscriptaddress,
 amsmath,amssymb,
 aps,prb,
]{revtex4-2}

\usepackage{graphicx}% Include figure files
\usepackage{dcolumn}% Align table columns on decimal point
\usepackage{bm}% bold math
\usepackage{color}
\usepackage{ulem}
\usepackage{xspace}
\usepackage{multirow}
\usepackage[unicode=true,colorlinks=true,linkcolor=blue,citecolor=blue,urlcolor=blue]{hyperref}
\usepackage{physics}
\usepackage{braket}
\usepackage{here}
\usepackage[utf8]{inputenc}

\usepackage{comment}
\usepackage{newtxtext}
\usepackage{newtxmath}
\usepackage{glossaries}

\setacronymstyle{long-short}
\newacronym{qha}{QHA}{quasiharmonic approximation}
\newacronym{ifc}{IFC}{interatomic force constant}
\newacronym{zsisa}{ZSISA}{zero static internal stress approximation}
\newacronym{v-z}{v-ZSISA}{volumetric ZSISA}
\newacronym{dft}{DFT}{density functional theory}
\newacronym{asr}{ASR}{acoustic sum rule}

\newcommand{\argmin}{\mathop{\rm argmin}\limits}

\DeclareUnicodeCharacter{2009}{\,} 
\definecolor{green}{rgb}{0,0.6,0.1}

\begin{document}

\preprint{APS/123-QED}

\title{Full optimization of quasiharmonic free energy with anharmonic lattice model: Application to thermal expansion and pyroelectricity of wurtzite GaN and ZnO}% Force line breaks with \\
%\thanks{A footnote to the article title}%
\author{Ryota Masuki}
\email{masuki-ryota774@g.ecc.u-tokyo.ac.jp}
\affiliation{
Department of Applied Physics, The University of Tokyo,7-3-1 Hongo, Bunkyo-ku, Tokyo 113-8656, Japan
}

\author{Takuya Nomoto}
\email{nomoto@ap.t.u-tokyo.ac.jp}
\affiliation{
Research Center for Advanced Science and Technology, The University of Tokyo,
4-6-1 Komaba Meguro-ku, Tokyo 153-8904, Japan
}

\author{Ryotaro Arita}
\email{arita@riken.jp}
\affiliation{
Research Center for Advanced Science and Technology, The University of Tokyo,
4-6-1 Komaba Meguro-ku, Tokyo 153-8904, Japan
}
\affiliation{ 
RIKEN Center for Emergent Matter Science, 2-1 Hirosawa, Wako, Saitama 351-0198, Japan 
}
\author{Terumasa Tadano}
\email{TADANO.Terumasa@nims.go.jp }
\affiliation{ 
CMSM, National Institute for Materials Science (NIMS), 1-2-1 Sengen, Tsukuba, Ibaraki 305-0047, Japan
}

%\collaboration{CLEO Collaboration}%\noaffiliation

\date{\today}% It is always \today, today,
             %  but any date may be explicitly specified

\begin{abstract}
We present a theory and a calculation scheme of structural optimization at finite temperatures within the quasiharmonic approximation (QHA). 
The theory is based on an efficient scheme of updating the interatomic force constants with the change of crystal structures, which we call the IFC renormalization. 
The cell shape and the atomic coordinates are treated equally and simultaneously optimized.
We apply the theory to the thermal expansion and the pyroelectricity of wurtzite GaN and ZnO, which accurately reproduces the experimentally observed behaviors.
Furthermore, we point out a general scheme to obtain correct $T$ dependence at the lowest order in constrained optimizations that reduce the number of effective degrees of freedom, which is helpful to perform efficient QHA calculations with little sacrificing accuracy. We show that the scheme works properly for GaN and ZnO by comparing with the optimization of all the degrees of freedom.
\end{abstract}

%\begin{description}
%\item[Usage]
%Secondary publications and information retrieval purposes.
%\item[Structure]
%You may use the \texttt{description} environment to structure your abstract;
%use the optional argument of the \verb+\item+ command to give the category of each item. 
%\end{description}
%\keywords{Suggested keywords}%Use showkeys class option if keyword
                              %display desired
\maketitle

%\tableofcontents

\section{Introduction}
\label{Sec_Introduction}

The thermophysical properties are among the most basic properties of solids, which play an important role in both fundamental science and various applications~\cite{10.3389/fchem.2018.00267, Liang2021, Miller2009, bowen2014pyroelectric, WANG2020105371, SURMENEV2021105442}. For its significant consequences, such as the thermal expansion and the pyroelectricity, it is essential to develop quantitative first-principles methods to understand and predict materials with desired properties.

The \gls{qha} is a widely used method~\cite{PhysRevB.71.205214, PhysRevB.61.8793, PhysRevLett.121.255901, PhysRevB.88.014303, PhysRevB.81.174301} that accurately computes the $T$-dependent crystal structure of weakly anharmonic solids~\cite{PhysRevB.92.064106, doi:10.1142/S0217984920500256, PhysRevB.105.064112}. In \gls{qha}, we neglect the anharmonic effect except for the crystal-structure dependence of the phonon frequencies $\{\hbar \omega_{\bm{k}\lambda}\}$ and approximate the free energy by the harmonic one~\cite{https://doi.org/10.1002/andp.19123441202, 10.2138/rmg.2010.71.3, doi:10.1063/1.5125779}.
The temperature-dependent crystal structure is obtained by minimizing the free energy with respect to the relevant structural degrees of freedom.
In the simple implementation, 
the phonon frequencies are calculated on a grid in the parameter space, and the free energy is fitted to calculate the temperature-dependent optimal parameters~\cite{Wang2008, PhysRevMaterials.5.085403, PhysRevLett.121.255901, doi:10.1063/1.5125779}.
This method works efficiently in optimizing a single degree of freedom, such as the lattice constant of a cubic material~\cite{doi:10.1063/1.5125779, TOGO20151, doi:10.1063/1.4928058}. However, the computational cost exponentially increases with the number of degrees of freedom $N_{\text{param}}$ because the phonon calculations must be performed on a multi-dimensional grid.

Several constrained optimization schemes have been proposed that reduce the number of effective degrees of freedom to perform calculations efficiently.
Using strain-dependent internal coordinates, which are determined to minimize the static potential energy, is the \gls{zsisa}~\cite{doi:10.1063/1.472684,  
PhysRevLett.120.207602,
doi:10.1063/5.0093376}.
\gls{zsisa} is correct for the $T$-dependent strain at the lowest order~\cite{doi:10.1063/1.472684}.
\gls{zsisa} combined with finite-temperature corrections of atomic shifts is used for calculating the pyroelectricity~\cite{PhysRevLett.120.207602}, which is actively studied recently~\cite{PhysRevB.93.081205, PhysRevApplied.12.034032, Liu_2019}.
In further approximation, the free energy is optimized with respect to the volume, while the other degrees of freedom are determined to minimize the static energy at fixed volumes~\cite{PhysRevB.81.174301, doi:10.1063/1.4896228, PhysRevB.76.064116, HUANG201684, NATH201682}. 
Based on these constrained optimizations, computational methods have also been devised to decrease computational costs further. The methods that use the Taylor expansion of the \gls{qha} free energy~\cite{doi:10.1021/acs.jctc.8b00460, PhysRevMaterials.6.043803} or the phonon frequencies~\cite{HUANG201684} and those focused on the irreducible representations of the symmetry groups are proposed~\cite{PhysRevB.106.014314}.
However, the internal coordinates are not optimized independently from the strain in these methods.

In this work, we develop a theory and a calculation scheme to optimize all the external and internal degrees of freedom within the quasiharmonic approximation. Our method is based on the \gls{ifc} renormalization, which efficiently updates the \glspl{ifc} using the anharmonic force constants~\cite{PhysRevB.106.224104, wallace1972thermodynamics}. Due to the compressive sensing method, which enables efficient extraction of the higher-order \glspl{ifc} from a small number of displacement-force data~\cite{PhysRevB.92.054301, PhysRevLett.113.185501, PhysRevB.100.184308}, the computational cost does not drastically increase for materials with many internal degrees of freedom. We apply the method to predict the thermal expansion and the pyroelectricity of wurtzite GaN and ZnO, for which we obtain reasonable agreements with the experimental results.

Furthermore, we prove a general theorem that provides an important guideline to efficiently get reliable results in constrained \gls{qha} optimizations.
The theorem is mathematically a straightforward generalization of a previous result on \gls{zsisa}~\cite{doi:10.1063/1.472684}, but it is helpful in designing constrained optimization schemes and clarifying their range of applicability.
Using the theorem, it is possible to get reasonable finite-temperature structures with $N_{\text{param}}$ separate one-dimensional optimizations instead of the grid search on $N_{\text{param}}$-dimensional parameter space, which decreases the computational cost from $O(N_{\text{s}}^{N_{\text{param}}})$ to $O(N_{\text{s}} N_{\text{param}})$, where $N_{\text{s}}$ is the number of sampling points of each parameter.
We implement \gls{zsisa} and several other constrained optimizations, whose results support the general statement.

\section{Theory}
\subsection{\Acrfull{qha}}
The anharmonic effect at each structure is neglected in the \gls{qha}. Thus, the \gls{qha} free energy of a crystal structure given by $X$ can be written as
\begin{align}
&
F_{\text{QHA}}(X, T) 
\nonumber
\\&= U_0(X) + \sum_{\bm{k}\lambda} \Bigl[ \frac{1}{2} \hbar \omega_{\bm{k}\lambda}(X) + k_{\mathrm{B}} T \log (1- e^{-\beta \hbar \omega_{\bm{k}\lambda}(X)}) \Bigr],
\end{align}
where $U_0(X)$ is the electronic ground state energy and $\omega_{\bm{k}\lambda}(X)$ is the $X$-dependent harmonic phonon frequency. $X$ consists of the external strain and the internal atomic positions. The crystal structure at finite temperature $T$ can be obtained by minimizing the \gls{qha} free energy as
\begin{align}
X(T) = \argmin_X F_{\text{QHA}}(X, T).
\end{align}
When combined with first-principles calculations, the most time-consuming part is the calculation of the structure dependence of the harmonic phonon frequencies $\omega_{\bm{k}\lambda}(X)$.

\subsection{\Acrfull{ifc} renormalization}
\label{subsec_IFC_renormalization}
We start from the Taylor expansion of the potential energy surface, which is introduced in Appendix~\ref{Appendix_Taylor_exp_of_PES}. The \gls{ifc} renormalization is a calculation method to update the set of \glspl{ifc} when the crystal structure is changed~\cite{PhysRevB.106.224104, wallace1972thermodynamics}. Since the new set of \glspl{ifc} are calculated from the \glspl{ifc} in the reference structure, there is no need to run additional electronic structure calculations at every step of the structure update, which makes the calculation significantly efficient.

The change of crystal structures can be described by the combination of the strain and the atomic displacements. 
We write the static atomic displacement in normal coordinate representation as
\begin{align}
  q^{(0)}_{\lambda} = \sum_{\alpha \mu} \epsilon_{\bm{0}\lambda,\alpha\mu} \sqrt{M_\alpha} u^{(0)}_{\alpha \mu},
\end{align}
where $u^{(0)}_{\alpha \mu}$ is the $\mu(=x,y,z)$ component of the static displacement of atom $\alpha$. $M_{\alpha}$ is the mass of atom $\alpha$ and $\epsilon_{\bm{0}\lambda, \alpha \mu}$ is the polarization vector of the mode $\lambda$ at $\Gamma$ point. 
$u^{(0)}_{\alpha \mu}$ is independent of the primitive cell $\bm{R}$ because we assume that the temperature-induced structural change is commensurate to $\Gamma$ point in the Brillouin zone. 

As for the strain, we use the displacement gradient tensor $u_{\mu \nu}$ as the basic variable, which is defined as
\begin{align}
    u_{\mu \nu} = \frac{\partial \widetilde{x}_{\mu}}{\partial x_{\nu}} - \delta_{\mu\nu}.
\end{align}
if the atom at $\bm{x}$ is moved to $\widetilde{\bm{x}}$ by the strain.
We restrict $u_{\mu \nu}$ to be symmetric to fix the rotational degrees of freedom.

The structural change described by the atomic displacements $q^{(0)}_{\lambda}$ corresponds to changing the center in the Taylor expansion of Eqs.~(\ref{eq_U_sum_of_Un}) and (\ref{eq_Un_sum_of_Phiu_Phiq}). As we have the polynomial form of the potential energy surface, which is determined by the \glspl{ifc} at the reference structure, it is possible to Taylor-expand again around the new structure. The expansion coefficient at the updated structure given by $q^{(0)}$ is written as  
\begin{align}
&
  \widetilde{\Phi}^{( q^{(0)})} (\bm{k}_1 \lambda_1, \cdots, \bm{k}_n \lambda_n)
  \nonumber \\=&
\sum_{m=0}^{\infty} \frac{1}{m!} \sum_{\{\rho\}}\widetilde{\Phi}^{( q^{(0)} =0)} (\bm{k}_1 \lambda_1, \cdots, \bm{k}_n \lambda_n, \bm{0}\rho_1, \cdots \bm{0}\rho_m) \nonumber \\
& \hspace{25mm} \times q^{(0)}_{\rho_1} \cdots q^{(0)}_{\rho_m}.
\label{EqRenormalizePhiByDisplace}
\end{align}
The derivation of the corresponding formula for the strain is more complicated. Although the strain is not included in the Taylor expansion of the potential energy surface [Eqs. (\ref{eq_U_sum_of_Un}) and (\ref{eq_Un_sum_of_Phiu_Phiq})], it is possible to recapture the strain as a set of static atomic displacements
\begin{align}
    u^{(0)}_{\bm{R}\alpha \mu} 
    &= 
    \sum_{\nu} u_{\mu \nu} (R_{\nu} + d_{\alpha\nu} )
    \sum_{\nu} u_{\mu \nu} R_{\alpha \nu},
    \label{Eq_uRau_uR}
\end{align}
where $\bm{d}_{\alpha}$ is the position of the atom $\alpha$ in the primitive cell. We define $\bm{R}_{\alpha} = \bm{R} + \bm{d}_{\alpha}$ for notational simplicity.
Thus, we can derive the \gls{ifc} renormalization in terms of strain as
\begin{align}
&
    \Phi^{(u_{\mu\nu})}_{\mu_1 \cdots \mu_n}(\bm{R}_1\alpha_1, \cdots, \bm{R}_n \alpha_n)
    \nonumber\\
    &= 
    \sum_{m=0}^{\infty} \frac{1}{m!} \nonumber \\
    &\times 
    \sum_{\{\bm{R}' \alpha' \mu' \nu'\}}
    \Phi^{(u_{\mu\nu}=0)}_{\mu_1 \cdots \mu_n \mu'_1 \cdots \mu'_m}(\bm{R}_1\alpha_1, \cdots, \bm{R}_n \alpha_n, \bm{R}'_1\alpha'_1, \cdots, \bm{R}'_m \alpha'_m)
    \nonumber
    \\&
    \hspace{25mm}
    \times 
    u_{\mu'_1 \nu'_1} R'_{1 \alpha'_1 \nu'_1}
    \cdots
    u_{\mu'_m \nu'_m} R'_{m \alpha'_m \nu'_m}.
    \label{EqRenormalizePhiByStrain}
\end{align}
See Ref.~\cite{PhysRevB.106.224104} for more detailed explanations.
Using Eqs. (\ref{EqRenormalizePhiByDisplace}) and (\ref{EqRenormalizePhiByStrain}), we can get the updated \glspl{ifc} for arbitrary strain and atomic displacements as long as the expansion from the reference structure is valid.
Hereafter, $\Phi$ and $\widetilde{\Phi}$ without notes in superscripts denote the renormalized \glspl{ifc} $\Phi^{(q^{(0)}, u_{\mu\nu})}$ and $\widetilde{\Phi}^{(q^{(0)}, u_{\mu\nu})}$, respectively, unless otherwise stated.

In the calculation, we truncate the Taylor expansion at the fourth order. As the \gls{ifc} renormalization by strain [Eq. (\ref{EqRenormalizePhiByStrain})] is written down in the real space, we first calculate them and Fourier-transform to the reciprocal space. The \gls{ifc} renormalization is performed in the order of $\widetilde{\Phi}^{(q^{(0)}=0, u_{\mu\nu}=0)} \to \widetilde{\Phi}^{(q^{(0)}=0, u_{\mu\nu})} \to \widetilde{\Phi}^{(q^{(0)}, u_{\mu\nu})}$. The details of the procedure is explained in Ref.~\cite{PhysRevB.106.224104}

Here, it should be noted that Eq. (\ref{EqRenormalizePhiByStrain}) is not directly applicable to the case $n=0$ because of the surface effect of the Born-von Karman supercell~\cite{wallace1972thermodynamics}, which we explain with an example in Appendix~\ref{Sec_Appendix_surfaceeffect_elasticconst}. As the solution for this problem is highly complicated, we expand the strain dependence of the potential energy surface as
\begin{align}
    \frac{1}{N}U_0^{(q^{(0)}=0, u_{\mu\nu})} 
    &= \frac{1}{2}\sum_{\mu_1 \nu_1, \mu_2 \nu_2} C_{\mu_1 \nu_1, \mu_2 \nu_2} \eta_{\mu_1 \nu_1} \eta_{\mu_2 \nu_2}
    \nonumber
    \\
    &+
    \frac{1}{6}\sum_{\mu_1 \nu_1, \mu_2 \nu_2, \mu_3 \nu_3} C_{\mu_1 \nu_1, \mu_2 \nu_2, \mu_3 \nu_3} \eta_{\mu_1 \nu_1} \eta_{\mu_2 \nu_2} \eta_{\mu_3 \nu_3}
    \nonumber
    \\&+ \cdots,
        \label{EqUpdateU0withStrain}
\end{align}
where $N$ is the number of primitive cells in the Born-von Karman supercell and
\begin{align}
    C_{\mu_1 \nu_1, \mu_2 \nu_2} = \frac{1}{N} \frac{\partial^2 U_0}{\partial \eta_{\mu_1 \nu_1} \partial \eta_{\mu_2 \nu_2}},
\end{align}
\begin{align}
    C_{\mu_1 \nu_1, \mu_2 \nu_2, \mu_3 \nu_3} = \frac{1}{N} \frac{\partial^2 U_0}{\partial \eta_{\mu_1 \nu_1} \partial \eta_{\mu_2 \nu_2} \partial \eta_{\mu_3 \nu_3}}, 
\end{align}
are the second and third-order elastic constants, which we define as the quantity per unit cell.
\begin{align}
    \eta_{\mu\nu} 
    &= 
    \frac{1}{2}\Bigl(\sum_{\mu'} (\delta_{\mu \mu'}+u_{\mu \mu'}) (\delta_{\nu\mu'}+u_{\nu \mu'}) - \delta_{\mu \nu}\Bigr)
    \\&=
    \frac{1}{2} 
    \Bigl(u_{\mu\nu} + u_{\nu \mu} + \sum_{\mu'} u_{\mu \mu'}u_{\nu \mu'}
    \Bigr).
\end{align}
is the strain tensor. The elastic constants are truncated at the third order in our calculation.

The \gls{ifc} renormalization in terms of atomic displacements [Eq. (\ref{EqRenormalizePhiByDisplace})] does not affect the fitting accuracy of the potential energy surface because it does not alter the potential landscape. However, the \gls{ifc} renormalization by strain [Eq. (\ref{EqRenormalizePhiByStrain})] is not necessarily precise because the information in a deformed cell is not provided in calculating the \glspl{ifc} in the reference structure. Thus, we estimate the coupling between the strain and the harmonic \glspl{ifc} 
\begin{align}
    \frac{\partial \Phi_{\mu_1 \mu_2}(\bm{R}_1\alpha_1, \bm{R}_2 \alpha_2)}{\partial u_{\mu\nu}},
\end{align}
using the finite displacement method with respect to the strain~\cite{PhysRevB.106.224104} to improve the accuracy of the method.

Additionally, the coupling between the first-order \glspl{ifc} and the strain
\begin{align}
    \frac{\partial \Phi(\bm{0}\lambda)}{\partial u_{\mu\nu}},
\end{align}
is also estimated using the finite displacement method of strain. This is because the acoustic sum rule of the first-order \glspl{ifc} is broken if the rotational invariance is not imposed on the harmonic \glspl{ifc}, which we explain in Appendix~\ref{Appendix_RotInv_ASR_firstorderIFC}. 
Since the rotational invariance imposes restrictions on \glspl{ifc} that the atomic forces calculated in the DFT supercell do not satisfy, it causes unreasonable shifts of the phonon frequencies. 
The frequency shifts depend on crystal symmetries, which makes the finite displacement estimation of $\frac{\partial \Phi_{\mu_1 \mu_2}(\bm{R}_1\alpha_1, \bm{R}_2 \alpha_2)}{\partial u_{\mu\nu}}$ difficult. Thus, we do not impose the rotational invariance on the harmonic \glspl{ifc} and calculate $\frac{\partial \Phi(\bm{0}\lambda)}{\partial u_{\mu\nu}}$ using the finite displacement method instead. The higher-order derivatives $\frac{\partial^2 \Phi(\bm{0}\lambda)}{\partial u_{\mu\nu} \partial u_{\mu'\nu'}}$, $\frac{\partial^3 \Phi(\bm{0}\lambda)}{\partial u_{\mu_1\nu_1} \partial u_{\mu_2\nu_2}  \partial u_{\mu_3\nu_3}}$ are set to zero because the rotational invariance of the higher-order \glspl{ifc} is required for them to satisfy the acoustic sum rule, which we also discuss in Appendix~\ref{Appendix_RotInv_ASR_firstorderIFC}. 

\subsection{Structural optimization within \gls{qha}}
Using the \gls{ifc} renormalization, the harmonic phonon dispersion and their derivatives can be calculated for updated crystal structures, which enables efficient minimization of the \gls{qha} free energy.
We begin with introducing a notation for the mode transformation. From here on, we distinguish the phonon modes in the reference structure and those in the updated structure. The former, which we write with greek letters without a bar (such as $\lambda$), is obtained by diagonalizing the dynamical matrix in the reference structure.
\begin{align}
&
\sum_{\beta \nu} 
\Bigl[\frac{1}{\sqrt{M_{\alpha} M_{\beta}}}
\sum_{\bm{R}} 
  \Phi^{(q^{(0)}=0, u_{\mu\nu}=0)}_{\mu\nu}(\bm{0}\alpha, \bm{R}\beta)
  e^{i\bm{k}\cdot \bm{R}}
  \Bigr]
  \epsilon_{\bm{k}\lambda,\beta \nu}
  \nonumber
  \\&= \omega_{\bm{k}\lambda}^2 \epsilon_{\bm{k}\lambda,\alpha \mu}.
\end{align}
These modes are fixed throughout the calculation, which serves as a reference frame.
The phonon modes in an updated structure, which we denote with a bar like $\bar{\lambda}$, diagonalize the dynamical matrix in the updated structure. We define the mode transformation matrix
\begin{align}
  C_{\bm{k}\lambda \bar{\lambda}} = \sum_{\alpha \mu} \epsilon^*_{\bm{k}\lambda, \alpha\mu} \epsilon_{\bm{k}\bar{\lambda}, \alpha\mu}.
\end{align}

Let us calculate the derivatives of the \gls{qha} free energy using the mode transformation.
Considering that the dynamical matrix is dependent on a parameter $s$, we can derive a formula
\begin{align}
  \frac{\partial (\omega^2_{\bm{k}\bar{\lambda}})}{\partial s}
  =
  \sum_{\lambda_1 \lambda_2} C^{*}_{\bm{k}\lambda_1 \bar{\lambda}} \frac{\partial \widetilde{\Phi}(-\bm{k}\lambda_1, \bm{k}\lambda_2)}{\partial s} C_{\bm{k}\lambda_2 \bar{\lambda}}.
\end{align}
Substituting $s = \widetilde{\Phi}(-\bm{k}\lambda_1, \bm{k}\lambda_2)$, we get
\begin{align}
\frac{\partial (\omega_{\bm{k}\bar{\lambda}})}{\partial \widetilde{\Phi}(-\bm{k}\lambda_1, \bm{k}\lambda_2)} = \frac{ C^{*}_{\bm{k}\lambda_1 \bar{\lambda}} C_{\bm{k}\lambda_2 \bar{\lambda}} }{2\omega_{\bm{k}\bar{\lambda}}}
\end{align}
Therefore, for a general structural degree of freedom $X_i$ that describes the atomic displacement $q^{(0)}_{\lambda}$ or the strain $u_{\mu \nu}$, the derivative of the \gls{qha} free energy can be calculated as
\begin{multline}
\frac{\partial F_{\text{QHA}}(X,T)}{\partial X_i} 
=
\frac{\partial U_0}{\partial X_i} 
\\
+
\sum_{\bm{k}\bar{\lambda}\lambda_1 \lambda_2} \frac{\hbar}{2} \frac{n_B(\hbar \omega_{\bm{k}\bar{\lambda}}) + 1/2}{\omega_{\bm{k}\bar{\lambda}}}  C^{*}_{\bm{k}\lambda_1 \bar{\lambda}} C_{\bm{k}\lambda_2 \bar{\lambda}} \frac{\partial \widetilde{\Phi}(-\bm{k}\lambda_1, \bm{k}\lambda_2)}{\partial X_i}
\label{eq_delFQHA_delX_general}
\end{multline}
The derivatives $\frac{\partial U_0}{\partial X_i}$ and $\frac{\partial \widetilde{\Phi}(-\bm{k}\lambda_1, \bm{k}\lambda_2)}{\partial X_i}$ can be obtained by differentiating Eqs. (\ref{EqRenormalizePhiByDisplace}), (\ref{EqRenormalizePhiByStrain}), and (\ref{EqUpdateU0withStrain}).

In our calculation, where the \glspl{ifc} are truncated at the fourth order and the elastic constants at the third order, the corresponding formulas are written as
\begin{widetext}
\begin{align}
\frac{1}{N} \frac{\partial U_0^{(q^{(0)}, u_{\mu\nu})}}{\partial q^{(0)}_{\lambda}} = \widetilde{\Phi}^{(q^{(0)}, u_{\mu\nu})}(\bm{0}\lambda),
\end{align}
\begin{align}
\frac{\partial \widetilde{\Phi}^{(q^{(0)}, u_{\mu\nu})}(\bm{k}_1\lambda_1, -\bm{k}_1 \lambda_2)}{\partial q^{(0)}_{\lambda}}
= 
\widetilde{\Phi}^{(q^{(0)}, u_{\mu\nu})}(\bm{k}_1\lambda_1, -\bm{k}_1 \lambda_2, \bm{0}\lambda),
\end{align}
\begin{align}
    \frac{1}{N}\frac{\partial U_0^{(q^{(0)}, u_{\mu\nu})}}{\partial u_{\mu\nu}}
    &=
    \sum_{\mu'\nu'}
    \frac{\partial \eta_{\mu'\nu'}}{\partial u_{\mu\nu}}
    \Bigl(
    \sum_{\mu_1 \nu_1} C_{\mu_1 \nu_1, \mu' \nu'} \eta_{\mu_1 \nu_1} 
    +
    \frac{1}{3}\sum_{\mu_1 \nu_1, \mu_2 \nu_2} C_{\mu_1 \nu_1, \mu_2 \nu_2, \mu' \nu'} \eta_{\mu_1 \nu_1} \eta_{\mu_2 \nu_2}
    \Bigr)
    \nonumber
    \\&
    +
    \sum_{m=1}^{3} \frac{1}{m!} 
    \sum_{\{\lambda\}} \frac{\partial \widetilde{\Phi}^{(q^{(0)}=0, u_{\mu\nu})}(\bm{0}\lambda_1, \cdots, \bm{0}\lambda_m)}{\partial u_{\mu\nu}} q^{(0)}_{\lambda_1} \cdots  q^{(0)}_{\lambda_m},
    \label{eq_delU0_delumunu}
\end{align}
\begin{align}
    \frac{\partial \widetilde{\Phi}^{(q^{(0)}, u_{\mu\nu})}(\bm{k}_1\lambda_1, -\bm{k}_1 \lambda_2)}{\partial u_{\mu\nu}}
    &=
    \frac{\partial \widetilde{\Phi}(\bm{k}_1\lambda_1, -\bm{k}_1 \lambda_2)}{\partial u_{\mu\nu}}
    +
    \sum_{\mu' \nu'} \frac{\partial^2 \widetilde{\Phi}(\bm{k}_1\lambda_1, -\bm{k}_1 \lambda_2)}{\partial u_{\mu\nu} \partial u_{\mu'\nu'}} u_{\mu'\nu'}
    \nonumber
    \\&+
    \sum_{\rho_1} \frac{\partial \widetilde{\Phi}(\bm{k}_1\lambda_1, -\bm{k}_1 \lambda_2, \bm{0}\rho_1)}{\partial u_{\mu\nu}} q^{(0)}_{\rho_1},
    \label{EqDelPhiDelUWithStrainDisplace}
\end{align}
\end{widetext}
for the internal coordinates and the strain respectively. The derivatives of the \glspl{ifc} in the RHS of Eq. (\ref{EqDelPhiDelUWithStrainDisplace}) are estimated at the reference structure ($q^{(0)}_{\lambda} = 0, u_{\mu \nu} = 0$).

Using the gradients of the free energy, we can simultaneously optimize all the internal and external degrees of freedom to minimize the \gls{qha} free energy. 
We denote the difference of the crystal structure from the optimum structure by $\delta q^{(0)}_{\lambda}$ and $\delta u_{\mu\nu}$. These quantities can be estimated by solving the linear equations
\begin{align}
\frac{1}{N}
    \frac{\partial F_{\text{QHA}}}{\partial q^{(0)}_{\lambda}} = \sum_{\lambda_1} \widetilde{\Phi}({\bm{0}\lambda, \bm{0}\lambda_1}) \delta q^{(0)}_{\lambda_1}
    \label{eq_lineareq_for_deltaq0}
\end{align}
\begin{align}
    \frac{1}{N}
    \frac{\partial F_{\text{QHA}}}{\partial u_{\mu\nu}} = \sum_{\mu_1 \nu_1} C_{\mu\nu, \mu_1 \nu_1} \delta u_{\mu_1 \nu_1},
    \label{EqLinearEqDeltaU}
\end{align}
where we approximate the Hessian of the QHA free energy by $\widetilde{\Phi}({\bm{0}\lambda, \bm{0}\lambda_1})$ and $C_{\mu\nu, \mu_1 \nu_1} $. 
We assume that $u_{\mu\nu}$ is symmetric to fix the rotational degrees of freedom, which is necessary to get a unique solution of Eq. (\ref{EqLinearEqDeltaU}).
The crystal structures are updated by
\begin{align}
    q^{(0)}_{\lambda} \leftarrow q^{(0)}_{\lambda} - \beta_{\text{mix,ion}} \delta q^{(0)}_{\lambda},
    \label{Eq_ion_update}
\end{align}
\begin{align}
    u_{\mu\nu} \leftarrow u_{\mu\nu} - \beta_{\text{mix,cell}} \delta u_{\mu\nu}.
    \label{Eq_cell_update}
\end{align}
The coefficients $\beta_{\text{mix,ion}}$ and $\beta_{\text{mix,cell}}$ are introduced for robust convergence of the calculation.
As for the constrained optimization methods such as \gls{zsisa}, we formulate different schemes of updating the crystal structure, which are described in detail in Appendix~\ref{Sec_Appendix_implement_ZSISA_vZSISA}.

\begin{figure}[h]
\vspace{0cm}
\begin{center}
\includegraphics[width=0.3\textwidth]{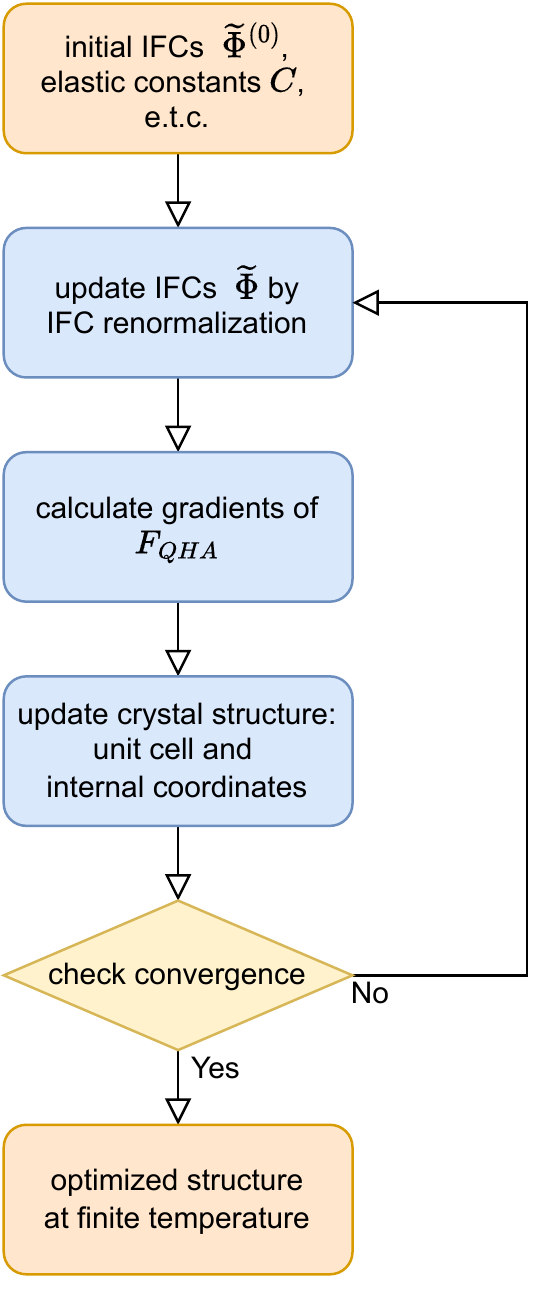}
\caption{
The calculation flow of the finite-temperature structural optimization within the quasiharmonic approximation combined with the \gls{ifc} renormalization.}
\label{Fig_calculation_scheme}
\end{center}
\end{figure}

From the above discussions, the calculation flow of the structural optimization based on \gls{ifc} renormalization and \gls{qha} is as follows, which we illustrate in Fig.~\ref{Fig_calculation_scheme}.
\begin{enumerate}
\item Input \glspl{ifc}, elastic constants, etc. at the reference structure. Define the initial structure.
\item Calculate the \glspl{ifc} in the current structure by \gls{ifc} renormalization.
\item Calculate gradients of the \gls{qha} free energy [Eqs. (\ref{eq_delFQHA_delX_general})-(\ref{EqDelPhiDelUWithStrainDisplace})].
\item Update the crystal structure [Eqs. (\ref{eq_lineareq_for_deltaq0})-(\ref{Eq_cell_update})].
\item Check convergence. If the convergence has yet to be achieved, go to 2. 
\end{enumerate}
We implement the theory to the ALAMODE package~\cite{Tadano_2014, PhysRevB.92.054301, PhysRevMaterials.3.033601}, which is an open-source software for anharmonic phonon calculation. The developed feature will be made public in its future release.

\subsection{General scheme of constrained optimizations correct at the lowest order}
\label{Sec_theorem_on_constrained_optimization}
Due to the high computational cost of optimizing all the degrees of freedom, numerous constrained optimization schemes have been proposed to decrease the number of effective degrees of freedom. 
\gls{zsisa} (\acrlong{zsisa}), which uses strain-dependent static internal coordinates~\cite{doi:10.1063/1.472684}, is a representative example.
In further approximation, the internal and deviatoric degrees of freedom are determined by minimizing the static energy~\cite{PhysRevB.81.174301, doi:10.1063/1.4896228, PhysRevB.76.064116, HUANG201684, NATH201682}, which we call \gls{v-z}. We illustrate \gls{zsisa} and \gls{v-z} with a schematic in Table~\ref{Table_ZSISA_vZSISA}.

Here, we show a general theorem on these constrained optimizations that reads \\\\
\textbf{Theorem.}
\textit{Consider optimizing the QHA free energy with respect to a set of structural degrees of freedom $\{X_i\}$. Then, if the other degrees of freedom $\{ \bar{X}_j \}$ are determined to minimize the static energy $U_0$ for given configurations of $\{X_i\}$, the obtained $T$ dependence of $\{X_i\}$ agrees at the lowest order with the result of the optimization of all the degrees of freedom (full optimization).}\\
{\ }

Mathematically, the theorem is just a straightforward corollary of the result in Ref.~\cite{doi:10.1063/1.472684}. However, we discuss it here because it will be a powerful guiding principle in designing an efficient and accurate constrained scheme of \gls{qha}. 
Before the proof of the theorem, we consider some of its applications, which we summarize in a list below.
\begin{itemize}
\item In \gls{zsisa}, $\{X_i\}$ represent the strain, and $\{ \bar{X}_j \}$ represent the internal coordinates. The theorem claims that $T$-dependence of the strain calculated by \gls{zsisa} is correct at the lowest order, which has been pointed out in Ref.~\cite{doi:10.1063/1.472684}.
\item In \gls{v-z}, $\{X_i\}$ represent the hydrostatic strain that causes volumetric expansion 
\begin{align}
u_{V,\mu\nu} \simeq
\left(
\begin{array}{ccc}
1 & 0 & 0\\
0 & 1 & 0 \\
0 & 0 & 1
\end{array}
\right),
\end{align}
while $\{ \bar{X}_j \}$ represent the deviatoric strain and the internal coordinates. According to the theorem, the volumetric expansion will be properly reproduced by \gls{v-z}.
\item $T$-dependence of an arbitrary degree of freedom $X_i$ can be calculated correctly at the lowest order if we relax all the other degrees of freedom in the static potential. This fact helps reduce the optimization of multiple degrees of freedom to the problem of separate optimization of each degree of freedom. Compared to the $N_{\text{param}}$-dimensional grid search of the computational cost of $O(N_{\text{s}}^{N_{\text{param}}})$, the computational cost of the separate one-dimensional optimization is decreased to $O(N_{\text{s}} N_{\text{param}})$, where $N_{\text{s}}$ is the number of sampling points of each parameter.
\\
e.g., consider the calculation of anisotropic expansion determined by two lattice constants, $a$ and $c$. The $T$-dependence of $a$ can be calculated by optimizing $c$ and the internal coordinate in the static potential. The $T$-dependence of $c$ can be calculated in a similar one-parameter optimization. The $T$-dependence of $c$ in the calculation of $a$ and that of $a$ in calculating $c$ should be disregarded.
\end{itemize}

It is worth mentioning that these constrained optimizations do not always reproduce the full optimization precisely because the higher-order effects can be nonnegligible in actual calculations. Nonetheless, in Secs.~\ref{subsec_resdis_ZSISA_vZSISA} and~\ref{subsec_resdis_const_opt_ac}, we discuss that the constrained optimization schemes based on the theorem give qualitatively accurate results more robustly than other schemes, once we determine the degrees of freedom to consider.

\begin{table*}[t]
  \caption{The schematic explanation of some different optimization schemes of \gls{qha}. Full optimization is the simultaneous optimization of all the degrees of freedom. In the table, \gls{qha} means that the degree of freedom is optimized at finite temperatures to minimize the \gls{qha} free energy, whereas static means that the degree of freedom is optimized in the static potential energy surface, which does not include the contribution of the lattice vibrations.}
  \begin{tabular}{|c|c|c|c|} 
    \hline & cell volume & deviatoric strain & atomic positions\\ \hline
    \hline
    & 
    \begin{minipage}{0.28 \textwidth}
      \centering
      \scalebox{0.2}{\includegraphics{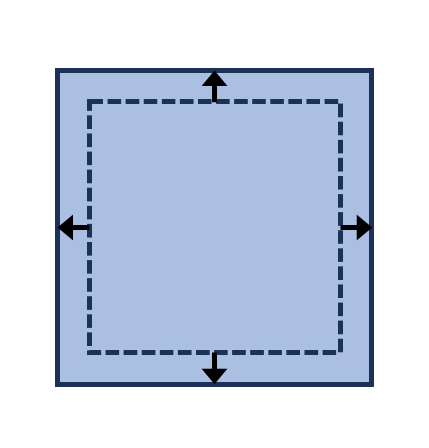}}
    \end{minipage} 
    &
    \begin{minipage}{0.28 \textwidth}
      \centering
      \scalebox{0.2}{\includegraphics{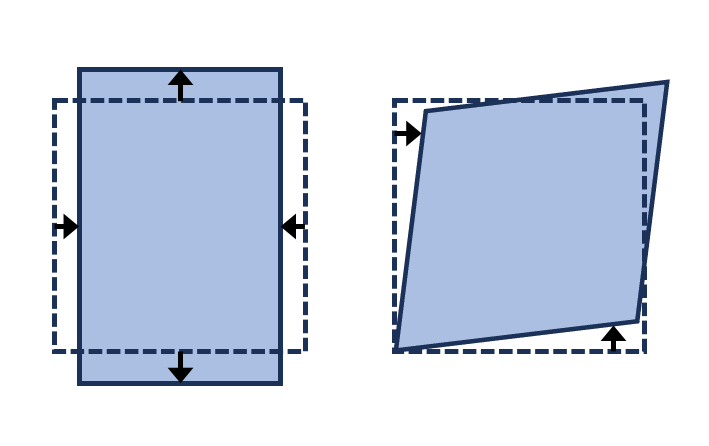}}
    \end{minipage}
    &
    \begin{minipage}{0.28 \textwidth}
      \centering
      \scalebox{0.2}{\includegraphics{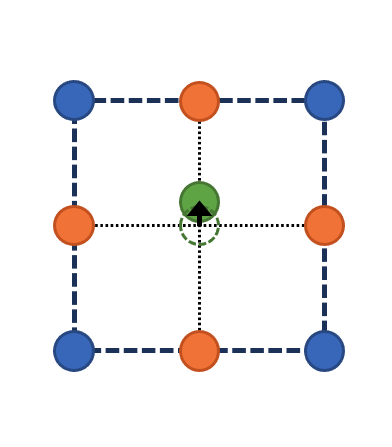}}
    \end{minipage} \\ \hline
    full optimization
    & QHA & QHA & QHA
    \\ \hline
    \gls{zsisa}
    & QHA & QHA & static
    \\ \hline
    \gls{v-z}
    & QHA & static & static
    \\ \hline
  \end{tabular}
  \label{Table_ZSISA_vZSISA}
\end{table*}

We move on to the proof of the theorem.
Since we assume that the reference structure is optimized in terms of the static potential $U_0$, the Taylor expansion of $U_0{(X, \bar{X})}$ is written as
\begin{align}
&
U_0{(X, \bar{X})} 
\nonumber\\
&= 
U_0{(X=\bar{X}=0)}
+
\frac{1}{2} \sum_{i_1  i_2} \frac{\partial^2 U_0}{\partial X_{i_1} \partial X_{i_2}} X_{i_1} X_{i_2} \nonumber\\&
+ \sum_{ij} \frac{\partial^2 U_0}{\partial X_{i} \partial \bar{X}_{j}} X_{i} \bar{X}_{j}
+\frac{1}{2} \sum_{j_1  j_2} \frac{\partial^2 U_0}{\partial \bar{X}_{j_1} \partial \bar{X}_{j_2}} \bar{X}_{j_1} \bar{X}_{j_2} 
+ \cdots .
\end{align}
The Taylor expansion of the \gls{qha} free energy is
\begin{align}
&
F_{\text{QHA}}(X, \bar{X}, T) 
\nonumber
\\&=
F_{\text{QHA}}(X = \bar{X} = 0, T)
+
\frac{1}{2} \sum_{i_1  i_2} \frac{\partial^2 U_0}{\partial X_{i_1} \partial X_{i_2}} X_{i_1} X_{i_2} \nonumber\\&
+ \sum_{ij} \frac{\partial^2 U_0}{\partial X_{i} \partial \bar{X}_{j}} X_{i} \bar{X}_{j}
+\frac{1}{2} \sum_{j_1  j_2} \frac{\partial^2 U_0}{\partial \bar{X}_{j_1} \partial \bar{X}_{j_2}} \bar{X}_{j_1} \bar{X}_{j_2} 
+ \cdots
\nonumber
\\&
+ \sum_i \frac{\partial F^{\text{vib}}_{\text{QHA}}}{\partial X_i} X_i
+ \sum_j \frac{\partial F^{\text{vib}}_{\text{QHA}}}{\partial \bar{X}_j} \bar{X}_j + \cdots .
\end{align}
Thus, in the lowest order approximation, the crystal structure that gives the minimum of the \gls{qha} free energy is calculated by solving
\begin{align}
\left(
\begin{array}{cc}
\dfrac{\partial^2 U_0}{\partial X \partial X} & \dfrac{\partial^2 U_0}{\partial X \partial \bar{X}} \\
\dfrac{\partial^2 U_0}{\partial \bar{X} \partial X } & \dfrac{\partial^2 U_0}{\partial \bar{X} \partial \bar{X}} \\
\end{array}
\right)
\left(
\begin{array}{c}
X \\
\bar{X} \\
\end{array}
\right)
= 
-
\left(
\begin{array}{c}
\dfrac{\partial F^{\text{vib}}_{\text{QHA}}}{\partial X} \\
\dfrac{\partial F^{\text{vib}}_{\text{QHA}}}{\partial \bar{X}} \\
\end{array}
\right).
\label{eq_lowestorder_fullQHA}
\end{align}
To eliminate $\bar{X}$ from the equation, we use 
\begin{align}
\bar{X}
=
-
\left(
\begin{array}{c}
\dfrac{\partial^2 U_0}{\partial \bar{X} \partial \bar{X}} 
\end{array}
\right)^{-1}
\left\{
\left(
\begin{array}{c}
\dfrac{\partial^2 U_0}{\partial X \partial \bar{X}}
\end{array}
\right)
X
+
\left(
\begin{array}{c}
\dfrac{\partial F^{\text{vib}}_{\text{QHA}}}{\partial \bar{X}} \\
\end{array}
\right)
\right\},
\end{align}
where we abbreviate the subscripts. The derivatives are estimated at $X = \bar{X} = 0$ in this section, except noted otherwise explicitly. Substituting to Eq. (\ref{eq_lowestorder_fullQHA}), we get
\begin{align}
&
\Bigl[
\left(
\begin{array}{c}
\dfrac{\partial^2 U_0}{\partial X \partial X}
\end{array}
\right)
-
\left(
\begin{array}{c}
\dfrac{\partial^2 U_0}{\partial X \partial \bar{X}}
\end{array}
\right)
\left(
\begin{array}{c}
\dfrac{\partial^2 U_0}{\partial \bar{X} \partial \bar{X}}
\end{array}
\right)^{-1}
\left(
\begin{array}{c}
\dfrac{\partial^2 U_0}{\partial \bar{X}\partial X }
\end{array}
\right)
\Bigr]
X
\nonumber
\\&
+ 
\left(
\begin{array}{c}
\dfrac{\partial F^{\text{vib}}_{\text{QHA}}}{\partial X} \\
\end{array}
\right)
-
\left(
\begin{array}{c}
\dfrac{\partial^2 U_0}{\partial X \partial \bar{X}}
\end{array}
\right)
\left(
\begin{array}{c}
\dfrac{\partial^2 U_0}{\partial \bar{X} \partial \bar{X}}
\end{array}
\right)^{-1}
\left(
\begin{array}{c}
\dfrac{\partial F^{\text{vib}}_{\text{QHA}}}{\partial \bar{X}} \\
\end{array}
\right)
=0
\label{eq_full_optimization_lowestorder_X}
\end{align}
as the equation for $X$.

Next, we consider the constrained optimization that $\bar{X}$ is determined to optimize $U_0$ for given configurations of $X$. 
In the lowest order,
\begin{align}
\left.
\left(
\begin{array}{c}
\dfrac{\partial U_0}{\partial \bar{X}}
\end{array}
\right)
\right|_{\bar{X} = \bar{X}(X)}
\simeq
\left(
\begin{array}{c}
\dfrac{\partial^2 U_0}{\partial \bar{X}\partial X }
\end{array}
\right)
X
+
\left(
\begin{array}{c}
\dfrac{\partial^2 U_0}{\partial \bar{X}\partial \bar{X} }
\end{array}
\right)
\bar{X}
=
0.
\end{align}
Hence, we get
\begin{align}
\bar{X}(X) = 
-
\left(
\begin{array}{c}
\dfrac{\partial^2 U_0}{\partial \bar{X}\partial \bar{X} }
\end{array}
\right)^{-1}
\left(
\begin{array}{c}
\dfrac{\partial^2 U_0}{\partial \bar{X}\partial X }
\end{array}
\right)
X.
\label{eq_XbarX}
\end{align}
Substituting to
\begin{align}
&
\left(
\begin{array}{c}
\dfrac{\partial F_{\text{QHA}}(X, \bar{X}(X), T)}{\partial X}
\end{array}
\right)
\nonumber
\\&=
\left(
\begin{array}{c}
\dfrac{\partial F_{\text{QHA}}}{\partial X}
\end{array}
\right)
+
\left(
\begin{array}{c}
\dfrac{\partial \bar{X}}{\partial X}
\end{array}
\right)
\left(
\begin{array}{c}
\dfrac{\partial F_{\text{QHA}}}{\partial \bar{X}}
\end{array}
\right)
\label{eq_delFqha_delX_XbarX}
\end{align}
we get
\begin{align}
&
\left(
\begin{array}{c}
\dfrac{\partial F_{\text{QHA}}(X, \bar{X}(X), T)}{\partial X}
\end{array}
\right)
\nonumber
\\&=
\Bigl[
\left(
\begin{array}{c}
\dfrac{\partial^2 U_0}{\partial X \partial X}
\end{array}
\right)
-
\left(
\begin{array}{c}
\dfrac{\partial^2 U_0}{\partial X \partial \bar{X}}
\end{array}
\right)
\left(
\begin{array}{c}
\dfrac{\partial^2 U_0}{\partial \bar{X} \partial \bar{X}}
\end{array}
\right)^{-1}
\left(
\begin{array}{c}
\dfrac{\partial^2 U_0}{\partial \bar{X}\partial X }
\end{array}
\right)
\Bigr]
X
\nonumber
\\&
+ 
\left(
\begin{array}{c}
\dfrac{\partial F^{\text{vib}}_{\text{QHA}}}{\partial X} \\
\end{array}
\right)
-
\left(
\begin{array}{c}
\dfrac{\partial^2 U_0}{\partial X \partial \bar{X}}
\end{array}
\right)
\left(
\begin{array}{c}
\dfrac{\partial^2 U_0}{\partial \bar{X} \partial \bar{X}}
\end{array}
\right)^{-1}
\left(
\begin{array}{c}
\dfrac{\partial F^{\text{vib}}_{\text{QHA}}}{\partial \bar{X}} \\
\end{array}
\right)
\label{delFQHA_delX_constrainedXbar}
\end{align}
Thus, the constrained optimization, which finds the solution of Eq. (\ref{delFQHA_delX_constrainedXbar}) = 0, is equivalent to the full optimization of Eq. (\ref{eq_full_optimization_lowestorder_X}) at the lowest order.

\subsection{Calculation of pyroelectricity}
\label{subsec_theory_calc_pyroelectricity}
We consider the effect of the static structural change for the $T$ dependence of the electric polarization $\bm{P}(T)$.
\begin{align}
P_{\mu}(T) = P_{\mu}(T=0) + 
\sum_{\alpha \nu} Z^*_{\alpha \mu \nu} u^{(0)}_{\alpha \nu}
+
\sum_{\mu_1 \nu_1}
d_{\mu, \mu_1 \nu_1} u_{\mu_1 \nu_1},
\label{eq_electric_polarization}
\end{align}
where $Z^*_{\alpha \mu \nu}$ is the Born effective charge, and $d_{\mu, \mu_1 \nu_1}$ is the ion-clamped piezoelectric tensor. We neglect the electron-phonon renormalization term, which originates from the thermal vibrations of the atoms~\cite{RevModPhys.17.245, PhysRevLett.35.1532, PhysRevLett.120.207602}.

The pyroelectricity is calculated by taking the temperature derivative of the spontaneous polarization.
\begin{align}
p_{\mu}(T) 
&= \frac{d P_{\mu}(T)}{d T}
\nonumber
\\&=
\sum_{\alpha \nu} Z^*_{\alpha \mu \nu} \frac{d u^{(0)}_{\alpha \nu}}{dT}
+
\sum_{\mu_1 \nu_1}
d_{\mu, \mu_1 \nu_1} \frac{d u_{\mu_1 \nu_1}}{dT}
\\&=
p_{\text{Born},\mu}(T) + p_{\text{piezo},\mu}(T).
\label{eq_def_pyroelectricity}
\end{align}

The pyroelectricity can also be split into the primary pyroelectricity $p^{(1)}$ and the secondary pyroelectricity $p^{(2)}$. The primary pyroelectricity is the clamped-lattice pyroelectricity, while the secondary pyroelectricity is the remaining part. Since $p_{\text{piezo}}$ is zero for fixed strains, $p_{\text{Born}}$ can be divided into the primary pyroelectricity and a part of the secondary pyroelectricity
\begin{align}
p_{\mu}(T) 
&=
p_{\text{Born},\mu}(T) + p_{\text{piezo},\mu}(T)
\nonumber
\\&=
p^{(1)}_{\mu}(T) + p^{(2)}_{\text{Born},\mu}(T) + p_{\text{piezo},\mu}(T).
\end{align}

\section{Simulation Details}
The developed method is applied to the thermal expansion and pyroelectricity of wurtzite GaN and ZnO.
In this section, we present the details of the calculation of these materials. Note that we use the same setting for both materials unless stated otherwise.

\subsection{Calculation of the interatomic force constants}
The lattice constants of the reference structures are determined by the structural optimization based on \gls{dft}; $a=$3.2183 \AA\ and $c=$5.2331 \AA\ for GaN, and $a=$3.2359 \AA\ and $c=$5.2247 \AA\ for ZnO.
The $4\times4\times2$ supercell, which contains 128 atoms, is employed for calculating the harmonic \glspl{ifc} of both GaN and ZnO. 
The Taylor expansion of the potential energy surface is truncated at the fourth order. 
For calculating the anharmonic \glspl{ifc}, the $3\times 3\times 2$ supercell containing 72 atoms is employed.
We generate 300 random configurations by uncorrelated random sampling from harmonic \glspl{ifc}~\cite{doi:10.1073/pnas.1707745115} at 500 K. The atomic forces are calculated by \gls{dft} calculations. The details of the \gls{dft} calculations are explained later in this section. The \glspl{ifc} are extracted from the obtained displacement-force data using adaptive LASSO implemented in the ALAMODE package~\cite{PhysRevB.92.054301}. The cutoff radii are set as 12 Bohr for cubic \glspl{ifc} and 8 Bohr for quartic \glspl{ifc}. The quartic \glspl{ifc} are restricted up to three-body terms.
We impose on the \glspl{ifc} the \gls{asr}, the permutation symmetry, and the space group symmetry considering the mirror images of the atoms in the supercell~\cite{PhysRevB.106.224104}.
The fitting error of the displacement-force data was 0.7696 \% for GaN and 2.1930 \% for ZnO, which indicates that the obtained set of \glspl{ifc} well captures the potential landscape.

The second and third-order elastic constants are calculated by fitting the strain-energy relation. The crystal symmetry is used to decrease the number of strain modes to calculate~\cite{PhysRevB.75.094105, LIAO2021107777, doi:10.1063/1.1714215}. For each strain mode, the ground state energy was calculated for 13 strained structures from $\eta = -0.03$ to $\eta = 0.03$ (See Ref.~\cite{LIAO2021107777} for the definition of $\eta$). The strain-energy relation was fitted by a cubic polynomial, whose coefficients are linear transformed to elastic constants.

The strain-IFC coupling constants $\frac{\partial \Phi_{\mu_1 \mu_2}(\bm{R}_1\alpha_1, \bm{R}_2 \alpha_2)}{\partial u_{\mu\nu}}$ and $\frac{\partial \Phi(\bm{0}\lambda)}{\partial u_{\mu\nu}}$ are determined by finite-difference method of first order. The harmonic \glspl{ifc} and the atomic forces are calculated for the six strain modes $u_{xx} = 0.005, u_{yy} = 0.005, u_{zz} = 0.005, u_{yz} = u_{zy} = 0.0025, u_{zx} = u_{xz} = 0.0025, u_{xy} = u_{yx} = 0.0025$. The other entries of the displacement gradient tensor $u_{\mu\nu}$ are zero in each strain mode.
Then, the coupling constants are obtained by dividing the differences from the results at the reference structure $u_{\mu \nu} = 0$.

In the \gls{qha} calculations, we use $8\times 8\times8$ $q$ mesh. We do not include nonanalytic correction in calculating the $T$-dependent crystal structures.

\subsection{Settings of the DFT calculations}
The \textit{Vienna ab initio simulation package} (VASP)~\cite{PhysRevB.54.11169} is employed for the electronic structure calculations. The PBEsol exchange-correlation functional~\cite{PhysRevLett.100.136406} and the PAW pseudopotentials~\cite{PhysRevB.50.17953, PhysRevB.59.1758} are used.
 The convergence criteria of the SCF loop is set to $10^{-8}$ eV, and accurate precision mode, which suppresses egg-box effects and errors, is used to calculate the forces accurately. The basis cutoff we use is 600 eV for both materials. We use a $4\times4\times4$ Monkhorst-Pack $k$-mesh for supercell calculations for both 4$\times$4$\times$2 and 3$\times$3$\times$2 supercells.
 The Born effective charges and the clamped-lattice piezoelectricity is calculated by density functional perturbation theory (DFPT)~\cite{PhysRevB.33.7017, PhysRevB.73.045112} in the reference structure.
 
\section{Results and Discussion}

\subsection{Finite-temperature structural optimization within \gls{qha}}
We apply the developed method to the thermal expansion and the pyroelectricity of wurtzite GaN and ZnO.
We first check the accuracy of the \gls{ifc} renormalization, which is shown to reproduce the results of \gls{dft} calculations correctly. Thus, the method can be regarded as a \gls{dft}-based first-principles calculation.
The result of the validations of the \gls{ifc} renormalization is summarized in Appendix~\ref{Appendix_test_IFCrenormalization}.

\begin{figure}[h]
\vspace{0cm}
\begin{center}
\includegraphics[width=0.48\textwidth]{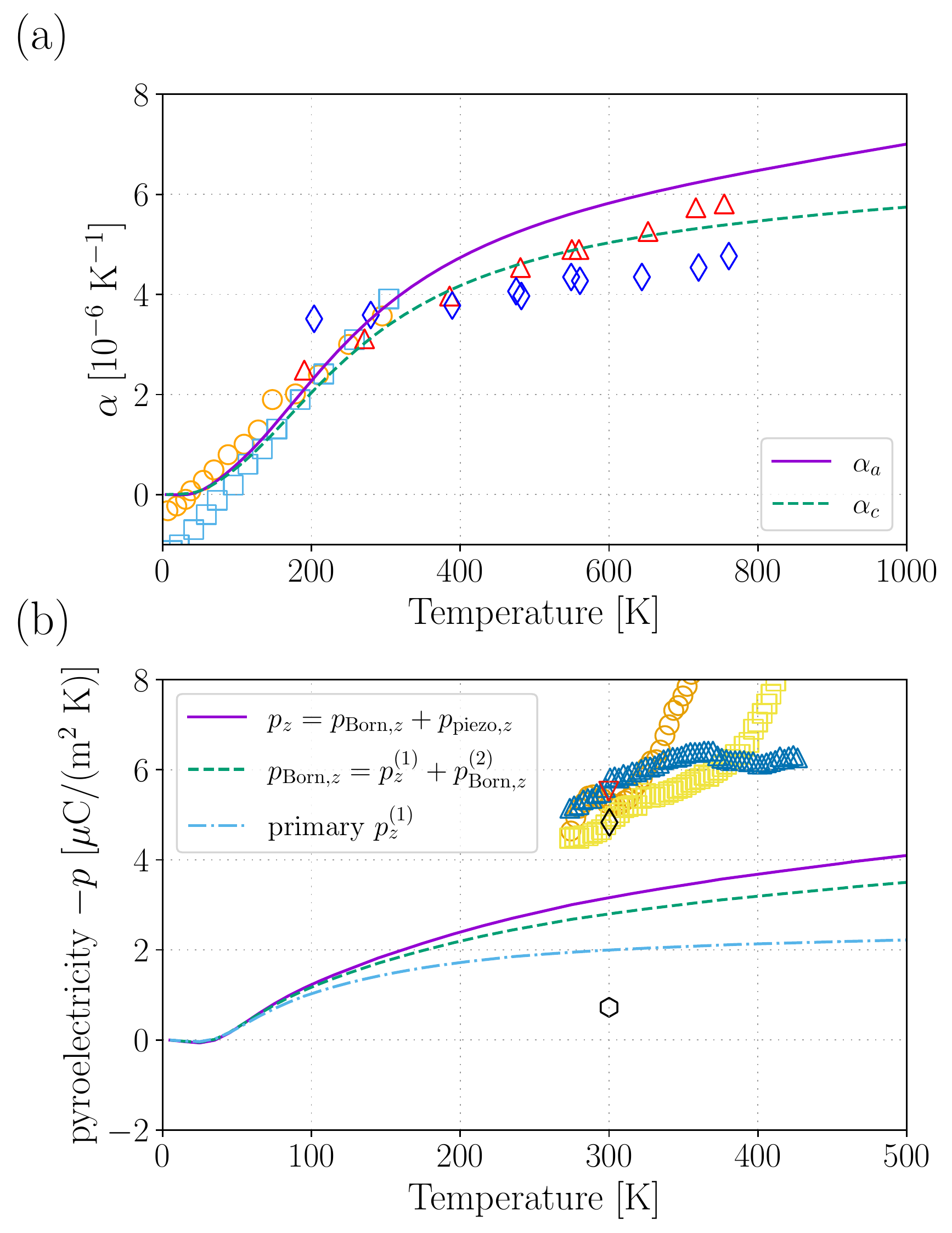}
\caption{
The thermal expansion and the pyroelectricity of GaN calculated by \gls{qha} combined with the \gls{ifc} renormalization. Both the internal coordinates and the strain are optimized to minimize the \gls{qha} free energy. 
(a) The thermal expansion coefficients of $a$ and $c$ axes ($\alpha_a = \frac{1}{a} \frac{da}{dT}$ and $\alpha_c = \frac{1}{c} \frac{dc}{dT}$ respectively). The experimental results are taken from Ref.~\cite{paszkowicz1999synchrotron} (orange circle for $\alpha_a$ and cyan square for $\alpha_c$) and Ref.~\cite{sheleg1976vestsi} (red triangle for $\alpha_a$ and blue diamond for $\alpha_c$).
(b) The purple line, green line, and cyan lines represent the total pyroelectricity, the Born term $p_{\text{Born},\mu} = \sum_{\alpha \nu} Z^*_{\alpha \mu \nu} \frac{d u^{(0)}_{\alpha \nu}}{dT}$, and the primary pyroelectricity $p^{(1)}_{\mu} = \sum_{\alpha \nu} Z^*_{\alpha \mu \nu} \Bigl(\frac{d u^{(0)}_{\alpha \nu}}{dT}\Bigr)_{\text{fixed\ cell}}$, which are defined in Sec.~\ref{subsec_theory_calc_pyroelectricity}.
The experimental results are taken from Ref.~\cite{doi:10.1063/1.4964265} (orange circle for the C-doped case, yellow square for the Fe-doped case, and blue triangle for the Mn-doped case), Ref.~\cite{981312} (red inverted triangle), Ref.~\cite{doi:10.1063/1.2716309} (black diamond), and Ref.~\cite{doi:10.1063/1.118027} (black hexagon).
}
\label{Fig_GaN_relaxall_aL_pyroelectricity}
\end{center}
\end{figure}

\begin{figure}[h]
\vspace{0cm}
\begin{center}
\includegraphics[width=0.48\textwidth]{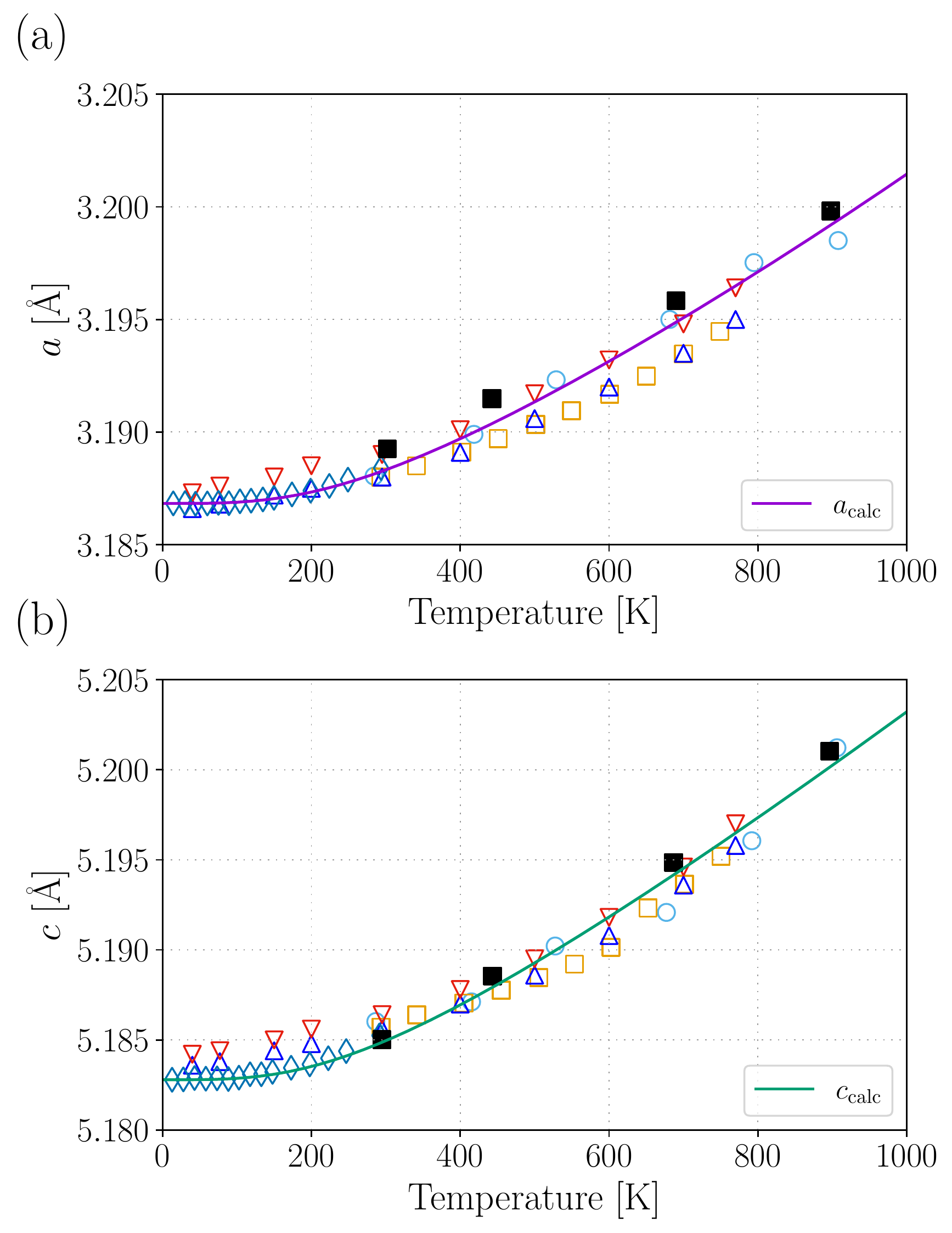}
\caption{
The temperature dependence of the lattice constants $a$ and $c$ of GaN calculated by \gls{qha} combined with the \gls{ifc} renormalization. Both the internal coordinates and the strain are optimized to minimize the \gls{qha} free energy. The calculation results are shifted by a constant to reproduce the experimental result at zero temperature. The experimental data are taken from Ref.~\cite{doi:10.1063/1.1652845} (cyan circle), Ref.~\cite{doi:10.1063/1.357273} (orange square for bulk), Ref.~\cite{leszczynski1996thermal} (blue triangle for bulk rough side and red inverted triangle for bulk smooth side), Ref.~\cite{reeber_wang_2000} (blue diamond), and Ref.~\cite{PhysRevB.72.085218} (black filled square).
}
\label{Fig_GaN_relaxall_ac}
\end{center}
\end{figure}

\begin{figure}[h]
\vspace{0cm}
\begin{center}
\includegraphics[width=0.48\textwidth]{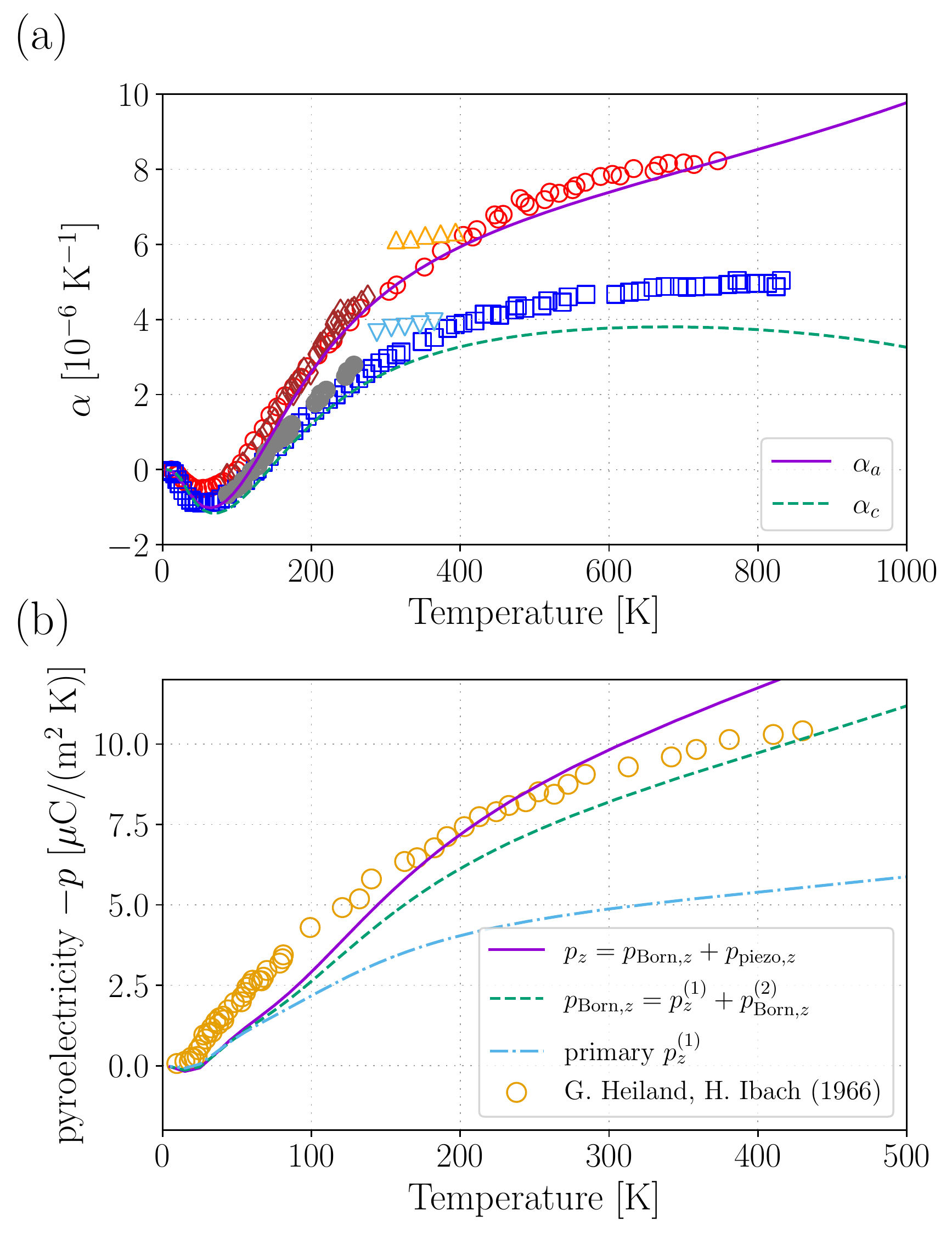}
\caption{
The thermal expansion and the pyroelectricity of ZnO calculated by \gls{qha} combined with the \gls{ifc} renormalization. Both the internal coordinates and the strain are optimized to minimize the \gls{qha} free energy.
(a) The thermal expansion coefficients of $a$ and $c$ axes ($\alpha_a = \frac{1}{a} \frac{da}{dT}$ and $\alpha_c = \frac{1}{c} \frac{dc}{dT}$ respectively).
The experimental data are taken from Ref.~\cite{https://doi.org/10.1002/pssb.19690330124}
(red circle for $\alpha_a$ and blue square for $\alpha_c$), Ref.~\cite{Khan:a05920} (orange triangle for $\alpha_a$ and cyan inverted triangle for $\alpha_c$), and Ref.~\cite{PhysRevB.4.1314} (brown diamond for $\alpha_a$ and gray filled circle for $\alpha_c$).
(b) The purple line, green line, and cyan lines represent the total pyroelectricity, the Born term $p_{\text{Born},\mu} = \sum_{\alpha \nu} Z^*_{\alpha \mu \nu} \frac{d u^{(0)}_{\alpha \nu}}{dT}$, and the primary pyroelectricity $p^{(1)}_{\mu} = \sum_{\alpha \nu} Z^*_{\alpha \mu \nu} \Bigl(\frac{d u^{(0)}_{\alpha \nu}}{dT}\Bigr)_{\text{fixed\ cell}}$, which are defined in Sec.~\ref{subsec_theory_calc_pyroelectricity}.
The experimental data is taken from Ref.~\cite{HEILAND1966353}.}
\label{Fig_ZnO_relaxall_aL_pyroelectricity}
\end{center}
\end{figure}

\begin{figure}[h]
\vspace{0cm}
\begin{center}
\includegraphics[width=0.48\textwidth]{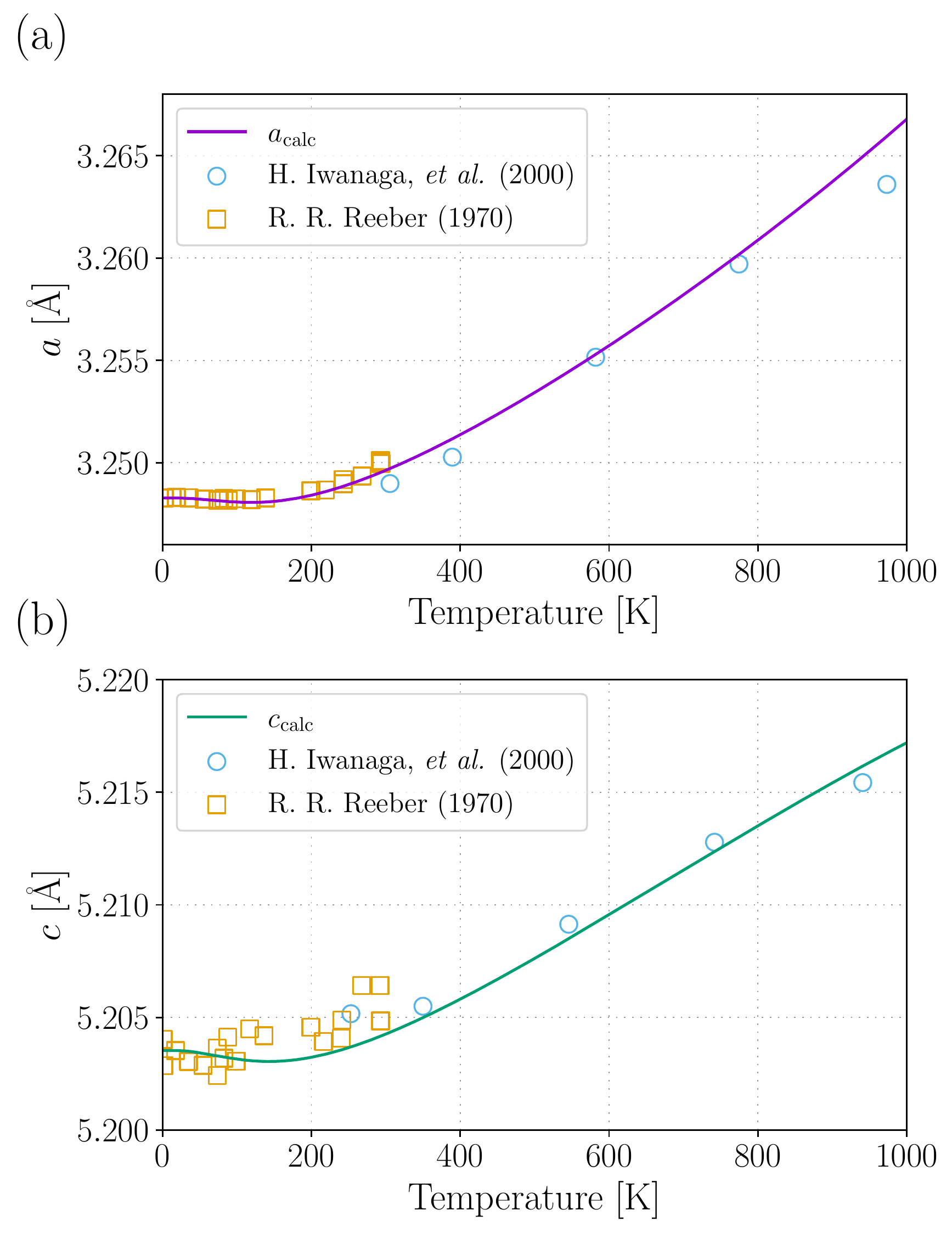}
\caption{
The temperature dependence of the lattice constants $a$ and $c$ of ZnO calculated by \gls{qha} combined with the \gls{ifc} renormalization. Both the internal coordinates and the strain are optimized to minimize the \gls{qha} free energy. The calculation results are shifted by a constant to reproduce the experimental result at zero temperature. The experimental data are taken from Ref.~\cite{Iwanaga2000} (cyan circle) and Ref.~\cite{doi:10.1063/1.1658600} (orange square).
}
\label{Fig_ZnO_relaxall_ac}
\end{center}
\end{figure}

Simultaneously optimizing both the internal coordinates and the strain within \gls{qha}, we get the calculation results shown in Figs.~\ref{Fig_GaN_relaxall_aL_pyroelectricity}--\ref{Fig_ZnO_relaxall_ac}.
As seen in the figures, the thermal expansion of both GaN and ZnO are quantitatively well reproduced with our method.
The thermal expansion is anisotropic, and the expansion coefficient of the lattice constant $a$ is larger than that of $c$.
This anisotropy is determined by a delicate interplay of internal and external degrees of freedom, which is accurately reproduced by the simultaneous optimization of all these degrees of freedom.

The calculation and experiment also show good agreement for the pyroelectricity as depicted in Figs.~\ref{Fig_GaN_relaxall_aL_pyroelectricity} (b) and~\ref{Fig_ZnO_relaxall_aL_pyroelectricity} (b).
The magnitude of the pyroelectricity is slightly underestimated for GaN. This can be because the experimental data are measured with thin films, not with bulk samples. Another possible reason is that the electron-phonon renormalization, which we neglect in this work, has a significant contribution, as proposed in Ref.~\cite{PhysRevLett.120.207602}.

\subsection{\gls{zsisa} and \gls{v-z}}
\label{subsec_resdis_ZSISA_vZSISA}
\begin{figure}[h]
\vspace{0cm}
\begin{center}
\includegraphics[width=0.48\textwidth]{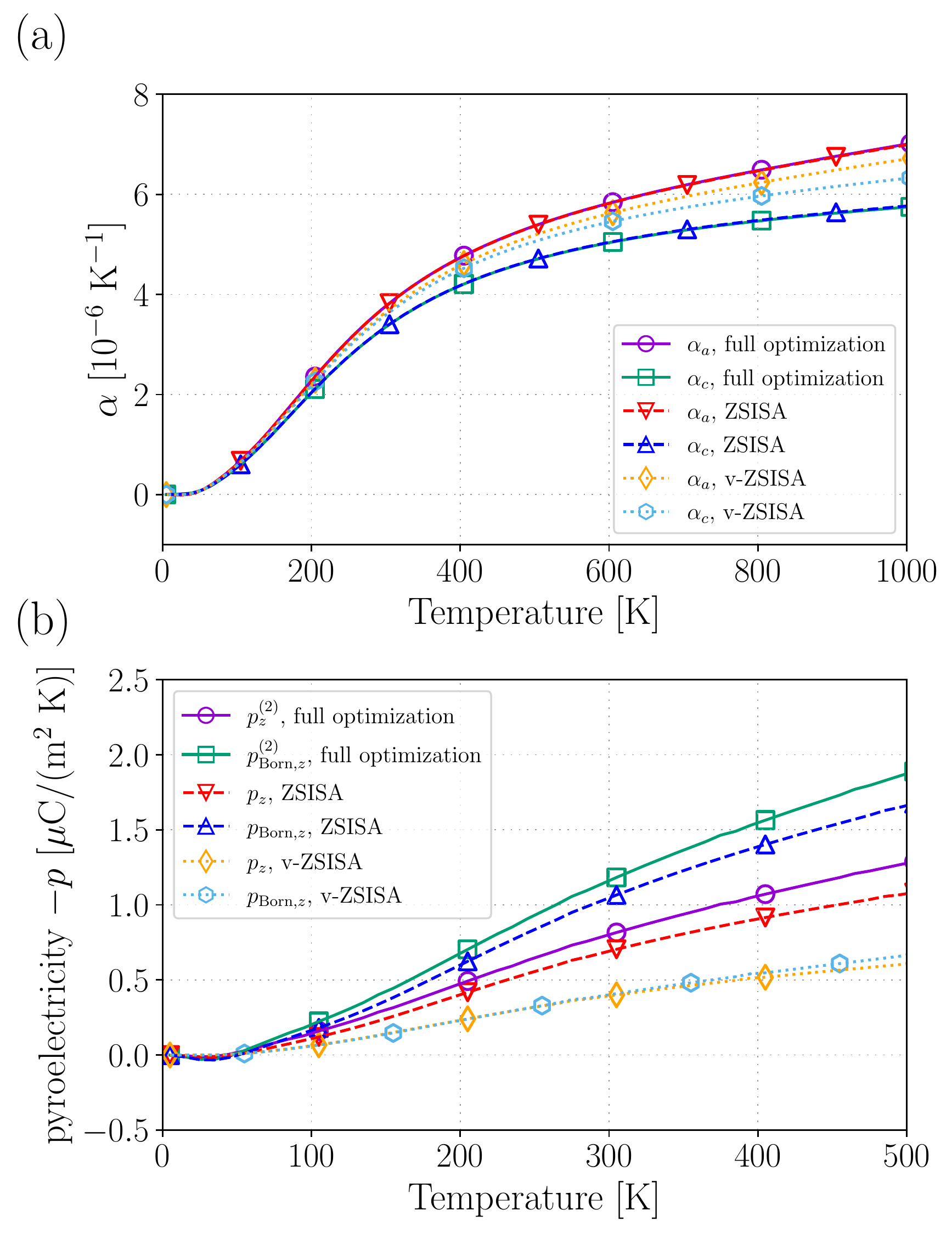}
\caption{
The thermal expansion and the pyroelectricity of GaN calculated by \gls{qha} combined with the \gls{ifc} renormalization. We compare the result of \gls{zsisa} and \gls{v-z} with the result of the simultaneous optimization of the internal coordinates and the strain (full optimization). 
(a) The thermal expansion coefficients of $a$ and $c$ axis ($\alpha_a = \frac{1}{a} \frac{da}{dT}$ and $\alpha_c = \frac{1}{c} \frac{dc}{dT}$ respectively). The full optimization results overlap with the \gls{zsisa} results. 
(b) The calculation results of the pyroelectricity. The secondary pyroelectricity $p^{(2)}_\mu = p^{(2)}_{\text{Born},\mu} + p_{\text{piezo},\mu} = \sum_{\alpha \nu} Z^*_{\alpha \mu \nu} \Bigl[\frac{d u^{(0)}_{\alpha \nu}}{dT} - 
\Bigl(
\frac{d u^{(0)}_{\alpha \nu}}{dT}
\Bigr)_{\text{fixed cell}}\Bigr] + \sum_{\mu_1 \nu_1}
d_{\mu, \mu_1 \nu_1} \frac{d u_{\mu_1 \nu_1}}{dT}
$
is plotted for the full optimization case, while the total pyroelectricity $p_z$ and the Born term $p^{(2)}_{\text{Born},z} = \sum_{\alpha \nu} Z^*_{\alpha \mu \nu} \frac{d u^{(0)}_{\alpha \nu}}{dT} $ are plotted for ZSISA and v-ZSISA. The different contributions to the pyroelectricity are defined in Sec.~\ref{subsec_theory_calc_pyroelectricity}.
}
\label{Fig_GaN_ZSISA_vZSISA_aL_pyroelectricity}
\end{center}
\end{figure}

\begin{figure}[h]
\vspace{0cm}
\begin{center}
\includegraphics[width=0.48\textwidth]{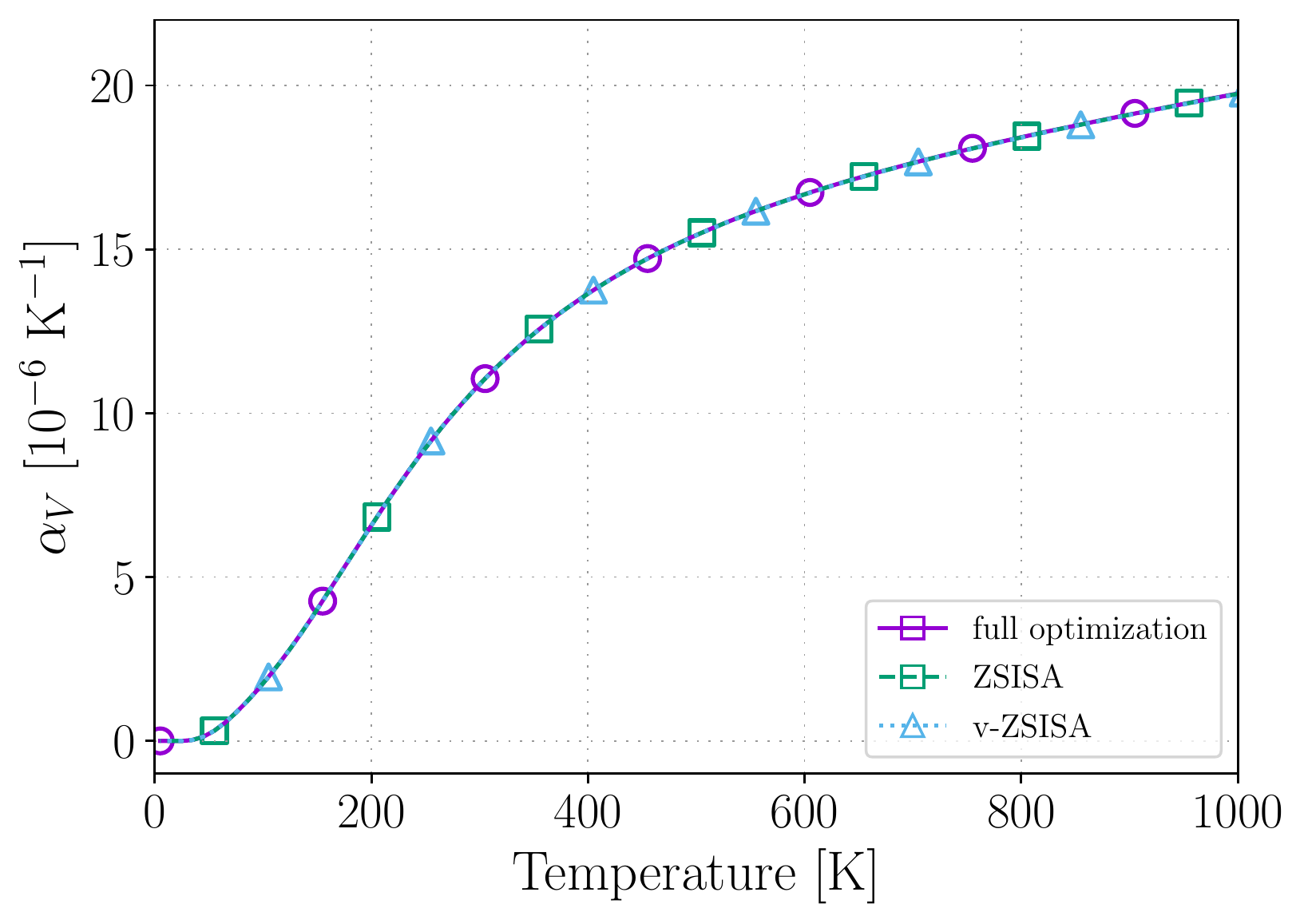}
\caption{
The volumetric thermal expansion coefficient $\alpha_V = \frac{1}{V}\frac{\partial V}{\partial T}$ of GaN calculated by \gls{qha} combined with \gls{ifc} renormalization. We compare the result of \gls{zsisa} and \gls{v-z} with the result of the simultaneous optimization of the internal coordinates and the strain (full optimization).
}
\label{Fig_GaN_ZSISA_vZSISA_aV}
\end{center}
\end{figure}

\begin{figure}[h]
\vspace{0cm}
\begin{center}
\includegraphics[width=0.48\textwidth]{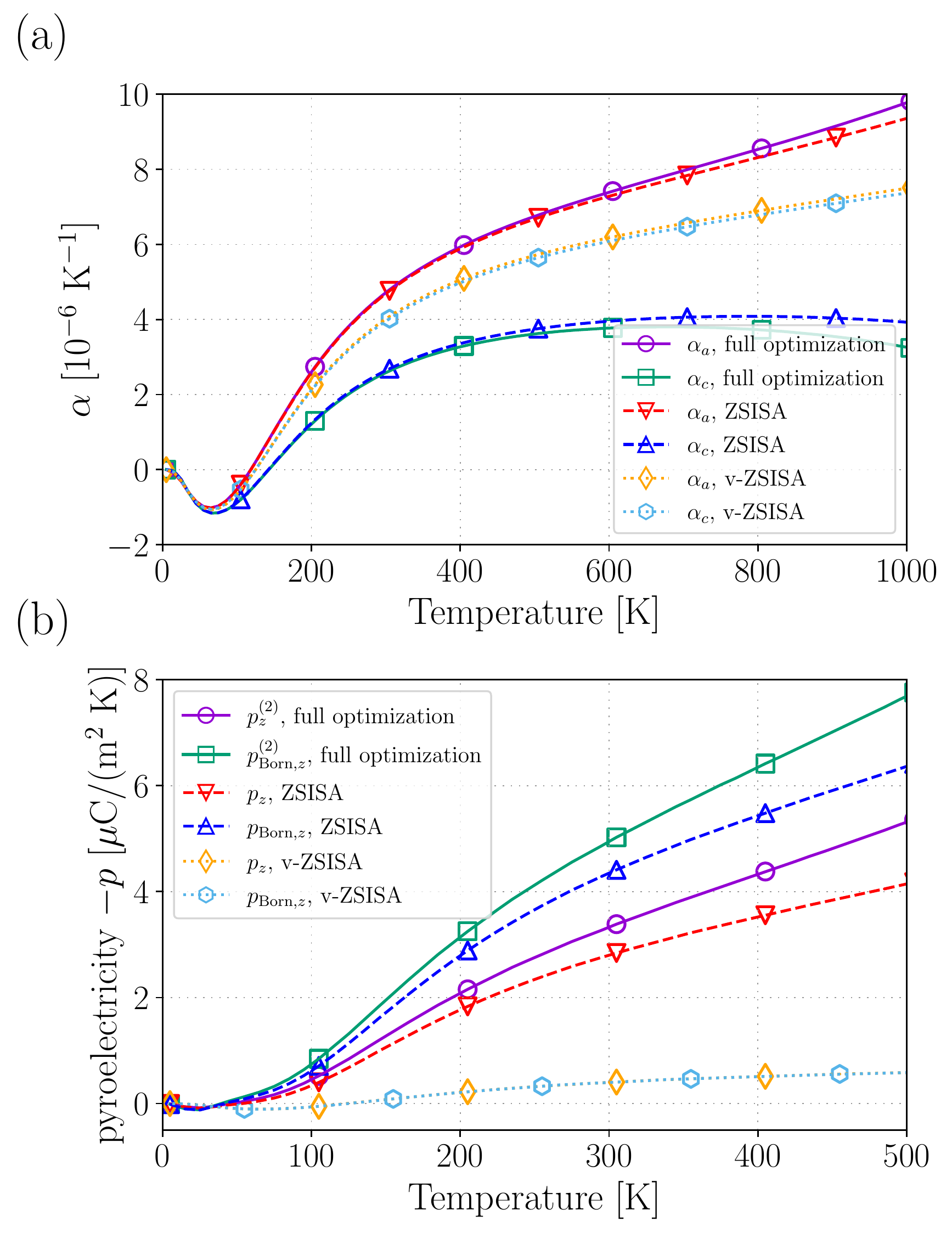}
\caption{
The thermal expansion and the pyroelectricity of ZnO calculated by QHA combined with the \gls{ifc} renormalization. We compare the result of ZSISA and v-ZSISA with the result of the simultaneous optimization of the internal coordinates and the strain (full optimization). 
(a) The thermal expansion coefficients of $a$ and $c$ axis ($\alpha_a = \frac{1}{a} \frac{da}{dT}$ and $\alpha_c = \frac{1}{c} \frac{dc}{dT}$ respectively). The full optimization results overlap with the ZSISA results. 
(b) The calculation results of the pyroelectricity. The secondary pyroelectricity $p^{(2)}_\mu = p^{(2)}_{\text{Born},\mu} + p_{\text{piezo},\mu} = \sum_{\alpha \nu} Z^*_{\alpha \mu \nu} \Bigl[\frac{d u^{(0)}_{\alpha \nu}}{dT} - 
\Bigl(
\frac{d u^{(0)}_{\alpha \nu}}{dT}
\Bigr)_{\text{fixed cell}}\Bigr] + \sum_{\mu_1 \nu_1}
d_{\mu, \mu_1 \nu_1} \frac{d u_{\mu_1 \nu_1}}{dT}
$
is plotted for the full optimization case, while the total pyroelectricity $p_z$ and the Born term $p^{(2)}_{\text{Born},z} = \sum_{\alpha \nu} Z^*_{\alpha \mu \nu} \frac{d u^{(0)}_{\alpha \nu}}{dT} $ are plotted for ZSISA and v-ZSISA. The different contributions to the pyroelectricity are defined in Sec.~\ref{subsec_theory_calc_pyroelectricity}.
}
\label{Fig_ZnO_ZSISA_vZSISA_aL_pyroelectricity}
\end{center}
\end{figure}

\begin{figure}[h]
\vspace{0cm}
\begin{center}
\includegraphics[width=0.48\textwidth]{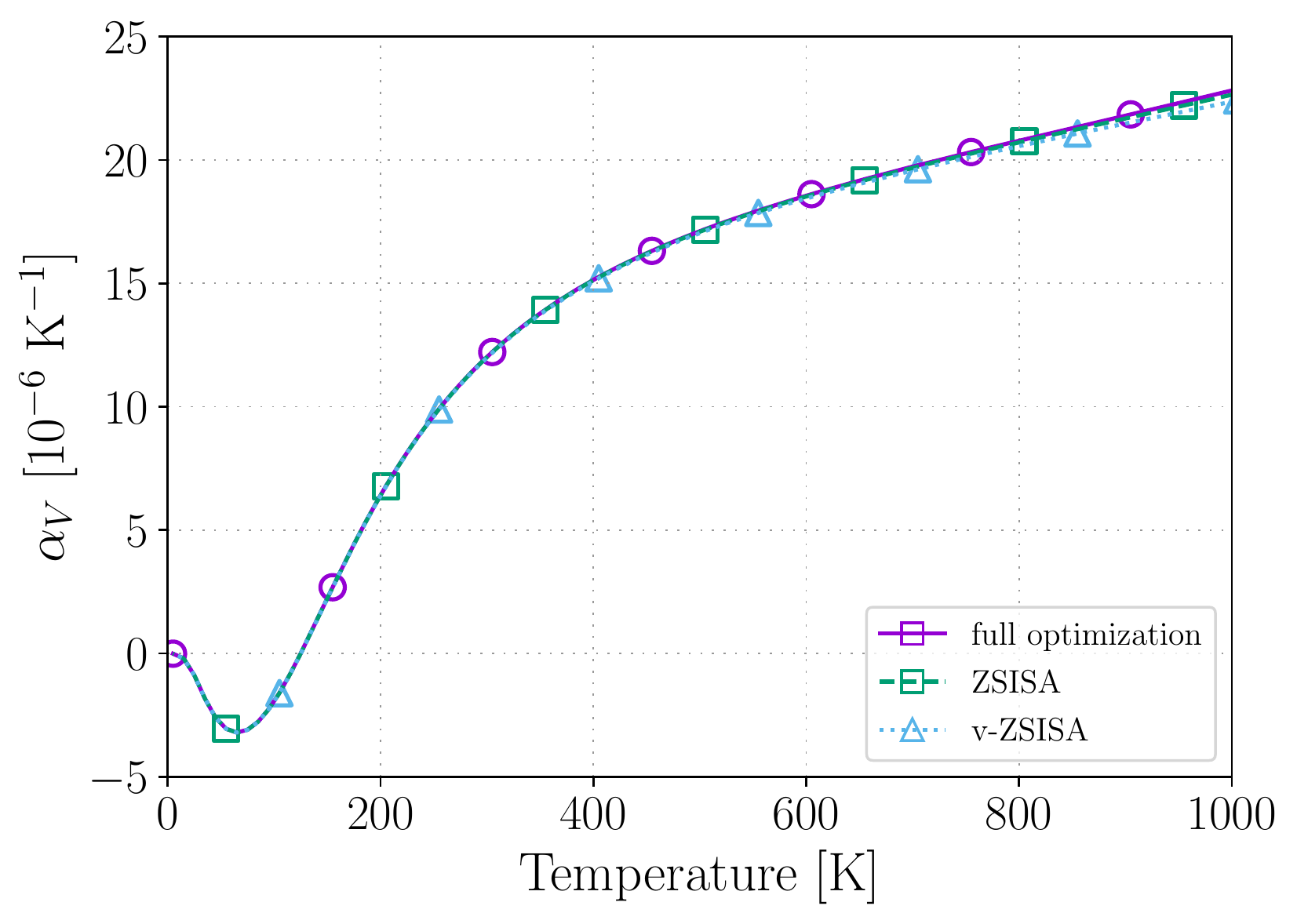}
\caption{
The volumetric thermal expansion coefficient $\alpha_V = \frac{1}{V}\frac{\partial V}{\partial T}$ of ZnO calculated by QHA combined with the \gls{ifc} renormalization. We compare the result of ZSISA and v-ZSISA with the result of the simultaneous optimization of the internal coordinates and the strain (full optimization).
}
\label{Fig_ZnO_ZSISA_vZSISA_aV}
\end{center}
\end{figure}

We perform the structural optimization using the \gls{ifc} renormalization in \gls{zsisa} and \gls{v-z}. The calculation results are shown in Figs.~\ref{Fig_GaN_ZSISA_vZSISA_aL_pyroelectricity}--\ref{Fig_ZnO_ZSISA_vZSISA_aV}.
According to Figs.~\ref{Fig_GaN_ZSISA_vZSISA_aL_pyroelectricity} (a) and ~\ref{Fig_ZnO_ZSISA_vZSISA_aL_pyroelectricity} (a), the thermal expansion coefficient calculated by \gls{zsisa} agrees well with the simultaneous optimization of all the degrees of freedom (full optimization). This is because \gls{zsisa} is 
correct at the lowest order for the $T$ dependence of the strain~\cite{doi:10.1063/1.472684}.
From Figs.~\ref{Fig_GaN_ZSISA_vZSISA_aL_pyroelectricity} (b) and ~\ref{Fig_ZnO_ZSISA_vZSISA_aL_pyroelectricity} (b), we can see that $T$-dependent pyroelectricity calculated by \gls{zsisa} well agrees with the secondary pyroelectricity in the full optimization, which is consistent with a previous calculation~\cite{LIU2018251}. 
As the internal coordinates are optimized at zero temperature in \gls{zsisa}, only the strain-induced secondary effects are taken into account. Some works add finite temperature effect of internal coordinates afterward as a correction~\cite{PhysRevLett.120.207602, LIU2018251}, which reproduces the full optimization results at the lowest order. 

We next look into the results of \gls{v-z}. As illustrated in  Figs.~\ref{Fig_GaN_ZSISA_vZSISA_aL_pyroelectricity} and~\ref{Fig_ZnO_ZSISA_vZSISA_aL_pyroelectricity}, \gls{v-z} significantly underestimates the anisotropy of the thermal expansion. As the $T$-dependent strain is not properly calculated, the secondary pyroelectricity is not correctly obtained either. However, as shown in Figs.~\ref{Fig_GaN_ZSISA_vZSISA_aV} and~\ref{Fig_ZnO_ZSISA_vZSISA_aV}, \gls{v-z} gives precise results for the volumetric thermal expansion coefficient. 
Here, we note that \gls{v-z} can be regarded as a special case of the constrained optimization scheme discussed in Sec.~\ref{Sec_theorem_on_constrained_optimization}. Because the volume of the unit cell is 
\begin{align}
v_{\text{cell}}(u_{\mu \nu}) 
&= 
v_{\text{cell}}(u_{\mu \nu}=0)  \times \det (I + u_{\mu \nu})
\nonumber
\\&\simeq
v_{\text{cell}}(u_{\mu \nu}=0) \times (1 + \Tr u_{\mu \nu}),
\end{align}
\gls{v-z} corresponds to optimizing the hydrostatic strain $\Tr u_{\mu \nu}$ or the cell volume at finite temperature while the other degrees of freedom are determined to minimize the DFT energy, 
which explains its success in calculating the volumetric expansion.
Hence, we elucidate the range of applicability of \gls{v-z}, that \gls{v-z} produces reliable results for the volumetric thermal expansion but not for the anisotropy and the internal coordinates.

\subsection{Constrained optimization of $a$ and $c$ axis}
\label{subsec_resdis_const_opt_ac}
We consider optimizing the $a$ axis and $c$ axis separately. 
Aside from the full optimization, we try 
three optimization schemes, which we explain for the case of calculating the $T$-dependence of $a$. 
The first one is a special case of the constrained optimization in Sec.~\ref{Sec_theorem_on_constrained_optimization}, which gives correct results for the considering degrees of freedom at the lowest order. In this method, we optimize the \gls{qha} free energy with respect to $a$ while we determine the $a$-dependence of $c$ and the internal coordinates by minimizing the static potential energy $U_0$ (constrained optimization for $a$). In the other two schemes, we fix $c$ at the value of the reference structure. The internal coordinates are also fixed in the second scheme (fixed $u^{(0)}_{\alpha\mu}$ and $c$), while they are
relaxed at the \gls{zsisa} level in the third one (ZSISA, fixed $c$).
We try similar calculation schemes for calculating the $T$-dependence of $c$ as well.

\begin{figure}[h]
\vspace{0cm}
\begin{center}
\includegraphics[width=0.48\textwidth]{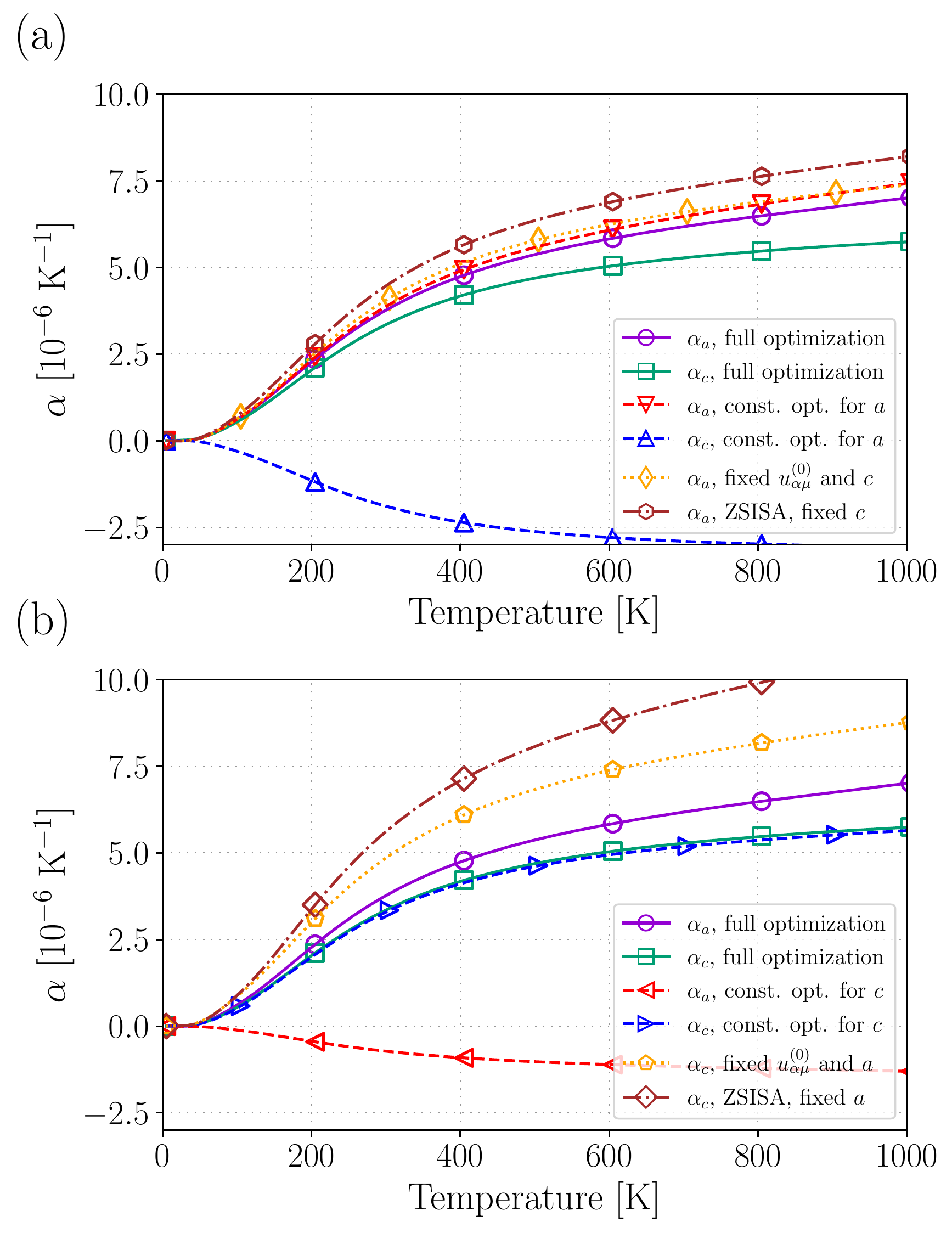}
\caption{
The thermal expansion coefficient of GaN calculated by \gls{qha} combined with the \gls{ifc} renormalization. $\alpha_a = \frac{1}{a} \frac{da}{dT}$ and $\alpha_c = \frac{1}{c} \frac{dc}{dT}$ are the thermal expansion coefficients of the $a$ and $c$ axis respectively. We compare the several schemes that separately calculate the temperature dependence of the lattice constants $a$ and $c$.
}
\label{GaN_ac_constrained_optimization}
\end{center}
\end{figure}

\begin{figure}[h]
\vspace{0cm}
\begin{center}
\includegraphics[width=0.48\textwidth]{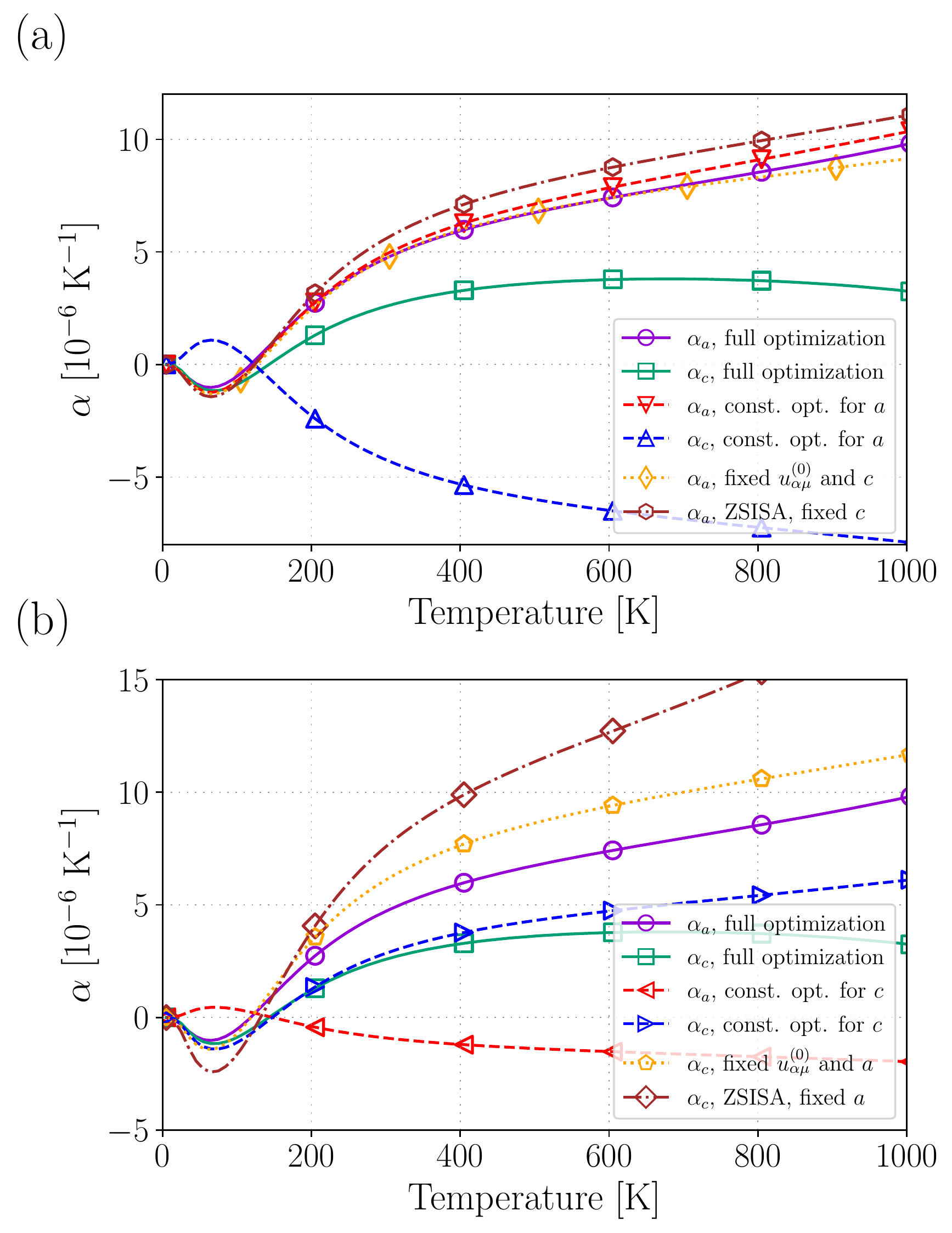}
\caption{
The thermal expansion coefficient of ZnO calculated by \gls{qha} combined with the \gls{ifc} renormalization. $\alpha_a = \frac{1}{a} \frac{da}{dT}$ and $\alpha_c = \frac{1}{c} \frac{dc}{dT}$ are the thermal expansion coefficients of the $a$ and $c$ axis respectively. We compare the several schemes that separately calculate the temperature dependence of the lattice constants $a$ and $c$.
}
\label{ZnO_ac_constrained_optimization}
\end{center}
\end{figure}

The calculation results are shown in Figs.~\ref{GaN_ac_constrained_optimization} and ~\ref{ZnO_ac_constrained_optimization}. 
As shown in Figs.~\ref{GaN_ac_constrained_optimization} (a) and ~\ref{ZnO_ac_constrained_optimization} (a), all the optimization schemes give similar results for $a$, which is close to the result obtained by simultaneous optimization of all degrees of freedom (full optimization).
Focusing on $\alpha_c$, the constrained optimization for $c$ well reproduces the results of the full optimization (Fig.~\ref{GaN_ac_constrained_optimization}(b)), albeit not precisely for ZnO (Fig.~\ref{ZnO_ac_constrained_optimization}(b)). The other methods that fix $a$ considerably overestimate the thermal expansion along the $c$ axis.
This reflects that the constrained optimization for $c$ is correct for calculating $T$ dependence of $c$ in the lowest order.
Note that the $T$-dependence of degrees of freedom that are relaxed in static potential (those in $\{\bar{X}_j\}$ in Sec.~\ref{Sec_theorem_on_constrained_optimization}) significantly deviates from the full optimization results. 
Therefore, the constrained optimization scheme discussed in Sec.~\ref{Sec_theorem_on_constrained_optimization} is useful to robustly get reasonable results by separately optimizing different degrees of freedom.

\section{Conclusions}
We formulate and develop a calculation method to simultaneously optimize all structural degrees of freedom, i.e., the strain and the internal coordinates, within the \acrfull{qha}. Our method is based on the Taylor expansion of the potential energy surface and the \gls{ifc} renormalization, which efficiently updates the \acrfullpl{ifc} with the change of crystal structures. We apply the method to the thermal expansion and the pyroelectricity of wurtzite GaN and ZnO, which shows good agreement with experiments.

Furthermore, we derive a general scheme of constrained optimization to obtain the correct $T$ dependence of considering structural degrees of freedom at the lowest order, in which we optimize all the other degrees of freedom in the static potential $U_0$. We perform calculations using several constrained optimization schemes, such as \gls{zsisa}, \gls{v-z}, and separate one-parameter optimization of $a$ and $c$ axis, whose results confirm the general scheme. 
Based on the general scheme, it is possible to reduce the optimization in the $N_{\text{param}}$-dimensional parameter space to $N_{\text{param}}$ separate one-parameter optimizations, which reduces the computational cost from $O(N_{\text{s}}^{N_{\text{param}}})$ to $O(N_{\text{s}} N_{\text{param}})$, where we denote the number of sampling points of each parameter as $N_{\text{s}}$.

\begin{acknowledgments}
This work was supported by JSPS KAKENHI Grant Number 21K03424 and 19H05825, Grant-in-Aid for JSPS Fellows (22J20892), and JST-PRESTO (JPMJPR20L7). 
\end{acknowledgments}

\appendix
\section{Taylor expansion of the potential energy surface}
\label{Appendix_Taylor_exp_of_PES}
We formulate the theory based on the Taylor expansion of the potential energy surface $\hat{U}$, which notation is introduced in this Appendix.
\begin{align}
\hat{U} = \sum_{n=0}^{\infty} \hat{U}_n,
\label{eq_U_sum_of_Un}
\end{align}
\begin{align}
&\hat{U}_n \nonumber \\
&=\frac{1}{n!} \sum_{\{\bm{R}\alpha \mu\}} \Phi_{\mu_1 \cdots \mu_n}(\bm{R}_1\alpha_1, \cdots, \bm{R}_n \alpha_n) \hat{u}_{\bm{R}_1 \alpha_1 \mu_1} \cdots \hat{u}_{\bm{R}_n \alpha_n \mu_n}
  \nonumber \\
&=\frac{1}{n!} \frac{1}{N^{n/2-1}} \sum_{\{\bm{k}\lambda\}} \delta_{\bm{k}_1 + \cdots + \bm{k}_n} \widetilde{\Phi} (\bm{k}_1 \lambda_1, \cdots, \bm{k}_n \lambda_n ) \hat{q}_{\bm{k_1} \lambda_1} \cdots \hat{q}_{\bm{k_n }\lambda_n}
  \label{eq_Un_sum_of_Phiu_Phiq},
\end{align}
where $\hat{u}_{\bm{R}\alpha \mu}$ is the $\mu(=x,y,z)$ component of the atomic displacement operator of atom $\alpha$ in the primitive cell at $\bm{R}$. $N$ is the number of primitive cells in the Born-von Karman supercell. The second line of Eq. (\ref{eq_Un_sum_of_Phiu_Phiq}) is the Fourier representation, which is defined by 
\begin{equation}
\hat{q}_{\bm{k}\lambda} = 
\frac{1}{\sqrt{N}}\sum_{\bm{R}\alpha \mu} 
e^{-i\bm{k}\cdot \bm{R}} 
\epsilon^*_{\bm{k}\lambda,\alpha\mu}\sqrt{M_\alpha} \hat{u}_{\bm{R}\alpha\mu},
\end{equation}
\begin{align}
&
\widetilde{\Phi}(\bm{k}_1 \lambda_1, \cdots, \bm{k}_n \lambda_n ) 
\nonumber\\
&=\frac{1}{N}\sum_{\{\bm{R}\alpha\mu\}} \Phi_{\mu_1 \cdots \mu_n}(\bm{R}_1\alpha_1, \cdots, \bm{R}_n \alpha_n) 
\nonumber\\& \times
\frac{\epsilon_{\bm{k}_1 \lambda_1,\alpha_1 \mu_1}}{\sqrt{M_{\alpha_1}}} e^{i\bm{k}_1\cdot \bm{R}_1} \cdots \frac{\epsilon_{\bm{k}_n \lambda_n,\alpha_n \mu_n}}{\sqrt{M_{\alpha_n}}} e^{i\bm{k}_n \cdot \bm{R}_n}
  \nonumber \\
&=\sum_{\{\alpha \mu\}} \frac{\epsilon_{\bm{k}_1 \lambda_1,\alpha_1 \mu_1}}{\sqrt{M_{\alpha_1}}}  \cdots \frac{\epsilon_{\bm{k}_n \lambda_n,\alpha_n \mu_n}}{\sqrt{M_{\alpha_n}}} 
\nonumber\\& \times 
  \sum_{\bm{R}_1\cdots \bm{R}_{n-1}} 
  \Phi_{\mu_1 \cdots \mu_n}(\bm{R}_1\alpha_1, \cdots, \bm{R}_{n-1} \alpha_{n-1}, \bm{0}\alpha_n)
  e^{i(\bm{k}_1\cdot \bm{R}_1+\cdots+\bm{k}_{n-1} \cdot \bm{R}_{n-1})}.\,
  \nonumber
\end{align}
where $M_{\alpha}$ is the mass of atom $\alpha$ and $\epsilon_{\bm{k}\lambda, \alpha \mu}$ is the polarization vector of the mode $\bm{k},\lambda$. The phonon modes are determined to diagonalize the harmonic dynamical matrix
\begin{align}
&
\sum_{\beta \nu} 
\Bigl[\frac{1}{\sqrt{M_{\alpha} M_{\beta}}}
\sum_{\bm{R}} 
  \Phi^{(q^{(0)}=0, u_{\mu\nu}=0)}_{\mu\nu}(\bm{0}\alpha, \bm{R}\beta)
  e^{i\bm{k}\cdot \bm{R}}
  \Bigr]
  \epsilon_{\bm{k}\lambda,\beta \nu}
  \nonumber
  \\&= \omega_{\bm{k}\lambda}^2 \epsilon_{\bm{k}\lambda,\alpha \mu}.
\end{align}
In the $n$-th order term $\hat{U}_n$, the expansion coefficients in real space $\Phi$ and in Fourier space $\widetilde{\Phi}$ are called the \acrfullpl{ifc}.

\section{Surface effects in the calculation of elastic constants using the \gls{ifc} renormalization}
\label{Sec_Appendix_surfaceeffect_elasticconst}
In Sec.~\ref{subsec_IFC_renormalization}, we derive the formula of the \gls{ifc} renormalization by the strain [Eq. (\ref{EqRenormalizePhiByStrain})], which is written down in the real-space representation.
However, this formula is not directly applicable to the renormalization of zeroth-order \gls{ifc} ($U_0$), which corresponds to the calculation of elastic constants. 
Since the finite-order \glspl{ifc} have fixed position arguments $\bm{R}_1\alpha_1, \cdots, \bm{R}_n \alpha_n$, the sum in the RHS of Eq.~(\ref{EqRenormalizePhiByStrain}) is restricted to a finite set of atoms around these $n$ atoms.
However, for the zeroth-order \gls{ifc}, the RHS of Eq.~(\ref{EqRenormalizePhiByStrain}) includes an infinite number of contributions from infinitely distant atoms, which is susceptible to the surface effects of the Born-von Karman supercell. 

In this Appendix, we explain this fact with a simple example. We consider a one-dimensional harmonic chain with nearest-neighbor interaction
\begin{align}
  \hat{U} = \frac{k}{2}\sum_j (\hat{u}_{j+1} - \hat{u}_j)^2 .
  \label{eq_harmonic_chain}
\end{align}
We omit $\mu = x,y,z$ and $\alpha$ because we consider a one-dimensional monatomic problem. We use an integer $i$ instead of $\bm{R}$ to describe the cell positions. The harmonic \glspl{ifc} in this problem are
\begin{align}
  &\Phi(j,j+1) = \Phi(j,j-1) = -k, \\
  &\Phi(j,j) = 2k.
\end{align}
Assume that the lattice constant $a$ of this harmonic chain is expanded to $a + \delta a$, and consider the change of the total energy $U_0$.
As the atomic displacement caused by this expansion is $u_j = j \delta a$ , we get
\begin{equation}
\delta U_0 
=\frac{k}{2}\sum_j (\hat{u}_{j+1} - \hat{u}_j)^2
=\frac{k}{2}\sum_j (\delta a)^2
=N \frac{k(\delta a)^2}{2}
\end{equation}
from Eq.~(\ref{eq_harmonic_chain}), 
which agrees with our intuition.
However, using Eq. (\ref{EqRenormalizePhiByStrain}), we get
\begin{align}
\delta U_0 
&=
\frac{1}{2} \sum_{jl} \Phi^{(\delta a = 0)}(j,l) u_j u_l
\nonumber
\\&=
\frac{1}{2} \sum_{j} u_j \Bigl[\sum_l\Phi^{(\delta a = 0)}(j,l)  u_l \Bigr]
\nonumber
\\&=
\frac{1}{2} \sum_{j} (\delta a)^2 j k [- (j+1) + 2j - (j-1)]
\nonumber
\\&=
0,
\end{align}
which is incorrect. This is because the transformation from the second to the third line is incorrect for the boundary atoms for which $l = j+1$ or $l = j-1$ do not exist. 
The contributions from the surface atoms are nonnegligible because their displacements caused by the strain are macroscopic.

From the above discussion, it is crucial to expand the potential using $(\bm{u}_{\bm{R}_i \alpha_i} - \bm{u}_{\bm{R}_j \alpha_j})$ in order to incorporate the surface effect correctly. This can be easily done in the one-dimensional case like Eq.~(\ref{eq_harmonic_chain}), but it is not straightforward in the higher-dimensional cases.
The \gls{ifc} renormalization from the harmonic and the cubic \glspl{ifc} are explained in Ref.~\cite{wallace1972thermodynamics}. Nevertheless, the treatment is highly complicated, and it is difficult to derive a general formula for arbitrary order. Therefore, we calculate elastic constants from DFT instead of using the \gls{ifc} renormalization. The computational cost is relatively small because the elastic constants can be calculated from the strain-energy relations of the primitive cell.

\section{Rotational invariance and \acrfull{asr} on the renormalized atomic forces}
\label{Appendix_RotInv_ASR_firstorderIFC}
\begin{widetext}
In \gls{ifc} renormalization by the strain, special care must be taken for the \acrfull{asr} of the first-order \glspl{ifc}. In $n$-th order \glspl{ifc} with $n \geq 2$, the renormalized \glspl{ifc} satisfy the ASR
\begin{align}
&
    \sum_{\bm{R}_n \alpha_n} \Phi_{\mu_1 \cdots \mu_{n-1} \mu_n}(\bm{0}\alpha_1, \cdots, \bm{R}_{n-1}\alpha_{n-1}, \bm{R}_n \alpha_n )
    = 0,
    \label{eq_ASR}
\end{align}
if the higher-order \glspl{ifc} of the reference structure satisfy the \gls{asr}. 
However, for the renormalized first-order \glspl{ifc} to satisfy the \gls{asr}, we show that the rotational invariance on the higher-order \glspl{ifc} must also be satisfied in the reference structure.
Note that we \sout{implicitly} assume that the \glspl{ifc} in the reference structure satisfy the \gls{asr} and the permutation symmetry, which assumption holds in our calculation. The space group symmetry is also imposed in the calculation, but it is not necessary for the discussion in this Appendix.
We show that\\
\\
\textbf{Proposition.} \textit{For $n\geq 2$, assume that the \gls{ifc} renormalization from the $(n-1)$-th order \glspl{ifc} to the first-order \glspl{ifc} satisfy the \gls{asr}. Then, if the rotational invariance between the $n$-th order and the $(n-1)$-th order \glspl{ifc} is satisfied, the \gls{ifc} renormalization from the $n$-th order \glspl{ifc} to the first-order \glspl{ifc} satisfy the \gls{asr}.}\\
\\
We start from the explanation of this statement.
The rotational invariance is the constraints on \glspl{ifc} which comes from the invariance of the total energy for rigid rotation of the whole system. The rotational invariance is a set of constraints that connects the $n$-th order and the $(n-1)$-th order \glspl{ifc}, which reads as follows.
\\
\\
\textit{The rotational invariance between the $n$-th order and the $(n-1)$-th order \glspl{ifc} is that Eq.~(\ref{eq_Rotinv_nn_1}) is symmetric under the exchange of $\mu$ and $\nu$.
\begin{align}
  \sum_{\bm{R}\alpha} \Phi_{\mu_1 \cdots \mu_n \mu } (\bm{R}_1 \alpha_1, \cdots, \bm{R}_n \alpha_n, \bm{R}\alpha) R_{\alpha \nu}
  +
  \sum_{i =1}^n \Phi_{\mu_i \to \mu} (\bm{R}_1 \alpha_1 \cdots \bm{R}_n \alpha_n) \delta_{\mu_i \nu},
  \label{eq_Rotinv_nn_1}
\end{align}
where $\mu_i \to \mu$ signifies $\mu_1 \cdots \mu_{i-1} \mu \mu_{i+1} \cdots \mu_n$.}\\
{\ }

The \gls{ifc} renormalization from the $n$-th order \glspl{ifc} to the first-order \glspl{ifc} by the strain is
\begin{align}
  \frac{\partial \Phi_{\mu} (\bm{0}\alpha_1)}{\partial u_{\mu_2 \mu_2} \cdots \partial u_{\mu_{n}\nu_{n}}}
  =
  \sum_{\{ \bm{R}\alpha \}} \Phi_{\mu_1 \cdots \mu_n} (\bm{0}\alpha_1 , \bm{R}_2 \alpha_2, \cdots, \bm{R}_n \alpha_n) R_{2\alpha_2 \nu_2} \cdots R_{n \alpha_n \nu_n}
\end{align}
Thus, the \gls{asr} on the \gls{ifc} renormalization from the $n$-th order \glspl{ifc} to the first-order \glspl{ifc} is
\begin{align}
  \sum_{\alpha_1}
  \sum_{\bm{R}_2\alpha_2 \cdots \bm{R}_n \alpha_n} \Phi_{\mu_1 \cdots \mu_n} (\bm{0}\alpha_1 , \bm{R}_2 \alpha_2, \cdots, \bm{R}_n \alpha_n) R_{2\alpha_2 \nu_2} \cdots R_{n \alpha_n \nu_n}
  = 0
  \label{Eq_nth_to_1st_ASR}
\end{align}
Let us now move onto the proof of the proposition. We first prove the following lemma.\\
\\
\textbf{Lemma 1.} \textit{The LHS of Eq.~(\ref{Eq_nth_to_1st_ASR}) is anti-symmetric under the exchange of $\mu_1 \leftrightarrow \mu_2$. }\\
\\
Starting from LHS of Eq.~(\ref{Eq_nth_to_1st_ASR}),
\begin{align}
  &
  \sum_{\alpha_1}
  \sum_{\bm{R}_2\alpha_2 \cdots \bm{R}_n \alpha_n} \Phi_{\mu_1 \cdots \mu_n} (\bm{0}\alpha_1 , \bm{R}_2 \alpha_2, \cdots, \bm{R}_n \alpha_n) R_{2\alpha_2 \nu_2} \cdots R_{n \alpha_n \nu_n}
  \nonumber
  \\&
  =
  \sum_{\alpha_1}
  \sum_{\bm{R}_2\alpha_2 \cdots \bm{R}_n \alpha_n} \Phi_{\mu_1 \cdots \mu_n} (\bm{R}_1\alpha_1 , \bm{R}_2 \alpha_2, \cdots, \bm{R}_n \alpha_n) (\bm{R}_{2\alpha_2} - \bm{R}_{1\alpha_1})_{\nu_2} \cdots (\bm{R}_{n \alpha_n} - \bm{R}_{1\alpha_1})_{\nu_n}
  \nonumber
  \\&=
  \sum_{\alpha_1}
  \sum_{\bm{R}_2\alpha_2 \cdots \bm{R}_n \alpha_n} \Phi_{\mu_1 \cdots \mu_n} (\bm{R}_1\alpha_1 , \bm{R}_2 \alpha_2, \cdots, \bm{R}_n \alpha_n) (\bm{R}_{2\alpha_2} - \bm{R}_{1\alpha_1})_{\nu_2} (\bm{R}_{3 \alpha_3} - \bm{R}_{2\alpha_2})_{\nu_n}\cdots (\bm{R}_{n \alpha_n} - \bm{R}_{2\alpha_2})_{\nu_n}
  \nonumber
  \\&=
  - \sum_{\alpha_1}
  \sum_{\bm{R}_2\alpha_2 \cdots \bm{R}_n \alpha_n} \Phi_{\mu_1 \leftrightarrow \mu_2} (\bm{R}_2\alpha_2 , \bm{R}_1 \alpha_1, \cdots, \bm{R}_n \alpha_n) (\bm{R}_{1\alpha_1} - \bm{R}_{2\alpha_2})_{\nu_2} (\bm{R}_{3 \alpha_3} - \bm{R}_{2\alpha_2})_{\nu_3}\cdots (\bm{R}_{n \alpha_n} - \bm{R}_{2\alpha_2})_{\nu_n}.
\end{align}
From the first line to the second line, we used the translational symmetry of the crystal lattice.
From the second to the third line, we use the acoustic sum rule on $i$-th atom for $i = 3, \cdots, n$.
Here, we note that $\bm{R}_1$ is not a dummy index but fixed somewhere in the crystal. Thus, the sum is restricted to a finite range where the atoms $\bm{R}_1$ and $\bm{R}_i$ interact. Although $\bm{R}_{i \alpha_i} - \bm{R}_{j \alpha_j}$ can be infinitely large for distant atoms, the sum can be considered as a finite sum of finite elements, which is extremely important to change the order of the summation.
We now fix $\bm{R}_2$ instead of $\bm{R}_1$, which is allowed due to the translational symmetry. Changing the names of the dummy indices and using the translational symmetry, we get
\begin{align}
  &
  \sum_{\alpha_1}
  \sum_{\bm{R}_2\alpha_2 \cdots \bm{R}_n \alpha_n} \Phi_{\mu_1 \cdots \mu_n} (\bm{0}\alpha_1 , \bm{R}_2 \alpha_2, \cdots, \bm{R}_n \alpha_n) R_{2\alpha_2 \nu_2} \cdots R_{n \alpha_n \nu_n}
  \nonumber
  \\&=
  -\sum_{\alpha_1}\sum_{\bm{R}_2\alpha_2 \cdots \bm{R}_n \alpha_n} \Phi_{\mu_1 \leftrightarrow \mu_2} (\bm{R}_1\alpha_1 , \bm{R}_2 \alpha_2, \cdots, \bm{R}_n \alpha_n) (\bm{R}_{2\alpha_2} - \bm{R}_{1\alpha_1})_{\nu_2} \cdots (\bm{R}_{n \alpha_n} - \bm{R}_{1\alpha_1})_{\nu_n}
  \nonumber
  \\&=
  -
  \sum_{\alpha_1}
   \sum_{\bm{R}_2\alpha_2 \cdots \bm{R}_n \alpha_n} \Phi_{\mu_1 \leftrightarrow \mu_2} (\bm{0}\alpha_1 , \bm{R}_2 \alpha_2, \cdots, \bm{R}_n \alpha_n) R_{2\alpha_2 \nu_2} \cdots R_{n \alpha_n \nu_n},
\end{align}
thus Lemma 1 has been proved.\\
\\
\textbf{Lemma 2.} \textit{Assume that the \gls{ifc} renormalization from the $(n-1)$-th order \glspl{ifc} to the first-order \glspl{ifc} satisfy the \gls{asr}, and the rotational invariance between the $n$-th order and the $(n-1)$-th order \glspl{ifc} is satisfied. Then the LHS of Eq.~(\ref{Eq_nth_to_1st_ASR}) is symmetric under the exchange of $\mu_2$ and $\nu_2$.}\\
\\
Again starting from the LHS of Eq.~(\ref{Eq_nth_to_1st_ASR}), 
\begin{align}
  &
  \sum_{\alpha_1}
  \sum_{\bm{R}_2\alpha_2 \cdots \bm{R}_n \alpha_n} \Phi_{\mu_1 \cdots \mu_n} (\bm{0}\alpha_1 , \bm{R}_2 \alpha_2, \cdots, \bm{R}_n \alpha_n) R_{2\alpha_2 \nu_2} \cdots R_{n \alpha_n \nu_n}
  \nonumber
  \\&=
  \sum_{\alpha_1}
  \sum_{\bm{R}_3\alpha_3 \cdots \bm{R}_n \alpha_n}
  R_{3\alpha_3 \nu_3} \cdots R_{n \alpha_n \nu_n}
  \Bigl[
    \sum_{\bm{R}_2 \alpha_2}
    \Phi_{\mu_1 \cdots \mu_n} (\bm{0}\alpha_1 , \bm{R}_2 \alpha_2, \cdots, \bm{R}_n \alpha_n) R_{2\alpha_2 \nu_2}
  \Bigr]
  \label{eq_lemma2_tochuushiki2}
\end{align}
Using the permutation symmetry of \glspl{ifc} and the rotational invariance between the $n$-th and the $(n-1)$-th order \glspl{ifc} [Eq.~(\ref{eq_Rotinv_nn_1})], we can show that Eq.~(\ref{eq_lemma2_tochuushiki1}) below is symmetric under the exchange of $\mu_2 \leftrightarrow \nu_2$
\begin{align}
  &
  \sum_{\alpha_1}
\sum_{\bm{R}_3\alpha_3 \cdots \bm{R}_n \alpha_n}
R_{3\alpha_3 \nu_3} \cdots R_{n \alpha_n \nu_n}
\Bigl[
  \sum_{\bm{R}_2 \alpha_2}
    \Phi_{\mu_1 \cdots \mu_n} (\bm{0}\alpha_1 , \bm{R}_2 \alpha_2, \cdots, \bm{R}_n \alpha_n) R_{2\alpha_2 \nu_2}
    \nonumber
    \\&
   +
    \sum_{i\neq 2} \delta_{\mu_i\nu_2} \Phi_{\mu_i \to \mu_2}(\bm{0}\alpha_1,\bm{R}_3 \alpha_3 , \cdots, \bm{R}_n \alpha_n)
\Bigr].
\label{eq_lemma2_tochuushiki1}
\end{align}
The second term in the square bracket vanishes when the summation is taken due to the \gls{asr} on the \gls{ifc} renormalization from the $(n-1)$-th order \glspl{ifc} to the first-order \glspl{ifc}.
Lemma 2 is derived by
comparing the RHS of Eqs.~(\ref{eq_lemma2_tochuushiki2}) and (\ref{eq_lemma2_tochuushiki1}).\\
\\
\textbf{Lemma 3.} \textit{Assume that the \gls{ifc} renormalization from the $(n-1)$-th order \glspl{ifc} to the first-order \glspl{ifc} satisfy the \gls{asr}, and the rotational invariance between the $n$-th order and the $n-1$-th order \glspl{ifc} is satisfied. Then the LHS of Eq.~(\ref{Eq_nth_to_1st_ASR}) is symmetric under the exchange of $\mu_1$ and $\nu_2$.}\\
\\
We show the last lemma for the proof of the proposition. We can use Lemma 1 and Lemma 2 from the assumption of Lemma 3. Thus, 
\begin{align}
(\mu_1 \mu_2, \nu_2)
=& - (\mu_2 \mu_1, \nu_2)\text{\ \ (Lemma 1)} \nonumber\\
=& - (\mu_2 \nu_2, \mu_1)\text{\ \ (Lemma 2)} \nonumber\\
=& (\nu_2 \mu_2, \mu_1)\text{\ \ (Lemma 1)},
\end{align}
where $(\mu_1 \mu_2, \nu_2)$ is a shorthand notation of the LHS of Eq.~(\ref{Eq_nth_to_1st_ASR}) which focuses on the permutation of the indices $\mu_1, \mu_2, \nu_2$.\\
\\
\textbf{proof of the proposition.}
Finally, we show the proof of the proposition. From Lemmas 2 and 3, we get
\begin{align}
(\mu_1 \mu_2, \nu_2)
=& (\nu_2 \mu_2, \mu_1) \text{\ \ (Lemma 3)} \nonumber \\
=& (\nu_2 \mu_1, \mu_2) \text{\ \ (Lemma 2)} \nonumber\\
=& (\mu_2 \mu_1, \nu_2) \text{\ \ (Lemma 3)}
\label{eq_proof_proposition_tochuushiki1}
\end{align}
On the other hand, Lemma 1 claims that
\begin{align}
(\mu_1 \mu_2, \nu_2)
= - (\mu_2 \mu_1, \nu_2) \text{\ \ (Lemma 1)}
\label{eq_proof_proposition_tochuushiki2}
\end{align}
Therefore, from Eqs.~(\ref{eq_proof_proposition_tochuushiki1}) and (\ref{eq_proof_proposition_tochuushiki2}), we get
\begin{align}
(\mu_1 \mu_2, \nu_2)
= 0,
\end{align}
which proves the proposition.
\end{widetext}

In the numerical calculation, we have confirmed the \gls{ifc} renormalization from the harmonic to the first-order \glspl{ifc} satisfies the \gls{asr} when we impose the rotational invariance on the harmonic \glspl{ifc}. On the other hand, we have checked that the \gls{ifc} renormalization to the first-order \glspl{ifc} from the higher-order \glspl{ifc} do not satisfy the \gls{asr} if we do not impose the rotational invariance. Therefore, it is numerically demonstrated that the \gls{asr} and the permutation symmetry alone are not sufficient for the \gls{asr} on the renormalized atomic forces to be satisfied.

\section{Implementations of \gls{zsisa} and \gls{v-z}}
\label{Sec_Appendix_implement_ZSISA_vZSISA}
The calculation of \gls{zsisa}, which fix the internal coordinates at the static positions in the potential energy surface, can be performed by fitting the strain-dependence of the free energy after relaxing the internal coordinate in the static potential. However, in our formalism combined with the \gls{ifc} renormalization, 
it is better to simultaneously optimize the internal and the external degrees of freedom to avoid the fitting error and to simplify the calculation scheme.
In \gls{v-z}, the complicated implementation of fixed-volume optimization will be a problem in calculating the volume-dependent \gls{v-z} free energy to curve-fit for minimization. 
In this Appendix, we explain that \gls{zsisa} and \gls{v-z} optimization can be performed by replacing the derivatives of \gls{qha} free energy in Eqs.~(\ref{eq_lineareq_for_deltaq0}) and (\ref{EqLinearEqDeltaU}) by appropriate functions.

We first explain the implementation of \gls{zsisa}. As the internal coordinates need to be relaxed to the static position of the potential $U_0$, we replace the RHS of Eq.~(\ref{eq_lineareq_for_deltaq0}) by 
\begin{align}
    \frac{\partial F_{\text{QHA}}}{\partial q^{(0)}_{\lambda}} 
    \to
    \frac{\partial F_{\text{ZSISA}}}{\partial q^{(0)}_{\lambda}} 
    = \frac{\partial U_0^{(q^{(0)}, u_{\mu \nu})}}{\partial q^{(0)}_{\lambda}} 
\end{align}
It should be emphasized that \gls{zsisa} is not formulated as a global minimization of a single function of internal coordinates $q^{(0)}$ and strain $u_{\mu\nu}$. Thus, $\frac{\partial F_{\text{ZSISA}}}{\partial q^{(0)}_{\lambda}} $ should not be interpreted as a derivative of a function $F_{\text{ZSISA}}$, but is used for notational simplicity.
The formula for the strain is similar to Eqs.~(\ref{eq_delFqha_delX_XbarX}) and (\ref{delFQHA_delX_constrainedXbar}) in Sec.~\ref{Sec_theorem_on_constrained_optimization}.
We define $\Bigl(\frac{\partial q^{(0)}_{\lambda}}{\partial u_{\mu \nu}} \Bigr)_{\text{ZSISA}}$ as the derivative in which $q^{(0)}$ is adjusted to the strain so that the atomic forces are invariant. This definition generalizes the derivative of the true strain dependence $q^{(0)}_{\lambda}(u_{\mu \nu})$ in \gls{zsisa} to arbitrary configurations of $q^{(0)}_{\lambda}$ and $u_{\mu \nu}$. The derivative can be calculated as
\begin{align}
\Bigl(\frac{\partial q^{(0)}_{\lambda}}{\partial u_{\mu \nu}} \Bigr)_{\text{ZSISA}}
&=
- \sum_{\lambda_1} (\widetilde{\Phi}_2^{-1})_{\lambda \lambda_1}
\Bigl(
\frac{\partial \widetilde{\Phi}(\bm{0}\lambda_1)}{\partial u_{\mu \nu}}
\Bigr),
\label{eq_delqdelu_ZSISA}
\end{align}
where $(\widetilde{\Phi}_2^{-1})$ is the inverse matrix of $\widetilde{\Phi}(\bm{0}\lambda_1, \bm{0} \lambda_2)$ in terms of the mode indices, which can be shown in a similar way to the derivation of Eq.~(\ref{eq_XbarX}) in Sec.~\ref{Sec_theorem_on_constrained_optimization}. 
The \glspl{ifc} and the derivatives in RHS of Eq.~(\ref{eq_delqdelu_ZSISA}) are estimated at the current structure with strain and atomic displacements.
The \gls{zsisa} derivative of the free energy is
\begin{align}
\frac{\partial F_{\text{ZSISA}}}{\partial u_{\mu \nu}} 
=
\frac{\partial F_{\text{QHA}}}{\partial u_{\mu \nu}} 
+
\sum_{\lambda}
\frac{\partial F_{\text{QHA}}}{\partial q^{(0)}_{\lambda}} 
\Bigl(\frac{\partial q^{(0)}_{\lambda}}{\partial u_{\mu \nu}} \Bigr)_{\text{ZSISA}},
\end{align}
with which we replace $\frac{\partial F_{\text{QHA}}}{\partial u_{\mu \nu}}$ in Eq.~(\ref{EqLinearEqDeltaU}).

In the calculation of \gls{v-z}, we separate the strain to the hydrostatic strain, which causes volumetric expansion, and the deviatoric strain. The mode of the hydrostatic strain $u_{V}$ is calculated as
\begin{align}
u_{V, \mu \nu} &\propto \frac{\partial \det (I+u)}{\partial u_{\mu \nu}}
\nonumber
\\&= (I+u)_{\mu+1,\nu+1} (I+u)_{\mu+2,\nu+2} - (I+u)_{\mu+1,\nu+2} (I+u)_{\mu+2,\nu+1},
\end{align}
where we use $x=0, y=1, z=2$ (mod 3) for notational simplicity.
We normalize $u_{V, \mu \nu}$ so that $\sum_{\mu \nu} |u_{V, \mu \nu}|^2 = 1$.
Here, we calculate the structural change $(\delta q^{(0)\text{v-ZSISA}}_{\lambda}, \delta u_{\mu \nu}^{\text{v-ZSISA}})$, in which the atomic forces and the deviatoric stress tensor are unaltered in the first order. These quantities can be obtained by solving the equation
\begin{align}
\left(
\begin{array}{cc}
\widetilde{\Phi}(\bm{0}\lambda , \bm{0}\lambda) & \dfrac{\partial \widetilde{\Phi}(\bm{0}\lambda)}{\partial u_{\mu \nu}} \\
\dfrac{\partial \widetilde{\Phi}(\bm{0}\lambda)}{\partial u_{\mu \nu}} & \widetilde{C}_{\mu_1 \nu_1, \mu_2 \nu_2}\\
\end{array}
\right)
\left(
\begin{array}{c}
\delta q^{(0)\text{v-ZSISA}}_{\lambda} \\
\delta u_{\mu \nu}^{\text{v-ZSISA}}\\
\end{array}
\right)
\propto
\left(
\begin{array}{c}
0 \\
 u_{V,\mu \nu}\\
\end{array}
\right),
\label{eq_deltaqdeltau_v-ZSISA}
\end{align}
where $\widetilde{C}_{\mu_1 \nu_1, \mu_2 \nu_2} = \frac{1}{N} \frac{\partial^2 U_0}{\partial u_{\mu_1 \nu_1} \partial u_{\mu_2 \nu_2}}$. The matrix elements in the LHS of Eq.~(\ref{eq_deltaqdeltau_v-ZSISA}) are IFC-renormalized by the strain and atomic displacements. 
We solve the equation assuming that the tensor $u_{\mu\nu}$ is symmetric to fix the rotational degrees of freedom.
We normalize the solution of Eq.~(\ref{eq_deltaqdeltau_v-ZSISA}) so that it satisfies
\begin{align}
\sum_{\mu \nu} u_{V, \mu \nu} \delta u_{\mu \nu}^{\text{v-ZSISA}} = 1.
\end{align}
Then, the \gls{v-z} derivative of the free energy in the direction of hydrostatic strain is 
\begin{align}
\frac{\partial F_{\text{v-ZSISA}}}{\partial u_{V}} 
&\propto 
\sum_{\mu \nu} \delta u_{\mu \nu}^{\text{v-ZSISA}} \frac{\partial F_{\text{QHA}}}{\partial u_{\mu \nu}} + \sum_{\lambda} \delta q^{(0)\text{v-ZSISA}}_{\lambda} \frac{\partial F_{\text{QHA}}}{\partial q^{(0)}_{\lambda}}
\nonumber
\\&=
\sum_{\mu \nu} \delta u_{\mu \nu}^{\text{v-ZSISA}} \frac{\partial F_{\text{ZSISA}}}{\partial u_{\mu \nu}} 
\end{align}
We denote the deviatoric strain modes, the modes perpendicular to $u_V$, as $u_i$. The \gls{v-z} derivative of the free energy in the direction of $u_i$ is 
\begin{align}
\frac{\partial F_{\text{v-ZSISA}}}{\partial u_i} 
=
\frac{\partial U_0^{(q^{(0)}, u_{\mu \nu})}}{\partial u_i},
\end{align}
since they should be relaxed in the static potential.
Transforming to the Cartesian representation, we get
\begin{align}
&
\frac{\partial F_{\text{v-ZSISA}}}{\partial u_{\mu \nu}} 
\nonumber
\\&
=
u_{V, \mu \nu} \sum_{\mu' \nu'}
\delta u_{\mu' \nu'}^{\text{v-ZSISA}} \frac{\partial F_{\text{ZSISA}}}{\partial u_{\mu' \nu'}}
\nonumber
\\&+ 
\Bigl(
\frac{\partial}{\partial u_{\mu \nu}} - u_{V, \mu \nu} \sum_{\mu' \nu'}
 u_{V,\mu' \nu'} \frac{\partial }{\partial u_{\mu' \nu'}}
\Bigr)
U_0^{(q^{(0)}, u_{\mu \nu})},
\end{align}
where the normalizations of $u_{V, \mu \nu}$ and $\delta u_{\mu' \nu'}^{\text{v-ZSISA}}$ are assumed. The \gls{v-z} derivative of the free energy in terms of the strain is 
\begin{align}
    \frac{\partial F_{\text{v-ZSISA}}}{\partial q^{(0)}_{\lambda}} 
    = \frac{\partial U_0^{(q^{(0)}, u_{\mu \nu})}}{\partial q^{(0)}_{\lambda}}.
\end{align}
The \gls{v-z} optimization can be performed by replacing the RHS of Eqs.~(\ref{eq_lineareq_for_deltaq0}) and (\ref{EqLinearEqDeltaU}) by $\frac{\partial F_{\text{v-ZSISA}}}{\partial q^{(0)}_{\lambda}} $ and $\frac{\partial F_{\text{v-ZSISA}}}{\partial u_{\mu \nu}}$ respectively.

\section{Test of \gls{ifc} renormalization}
\label{Appendix_test_IFCrenormalization}
\begin{figure}[h]
\vspace{0cm}
\begin{center}
\includegraphics[width=0.48\textwidth]{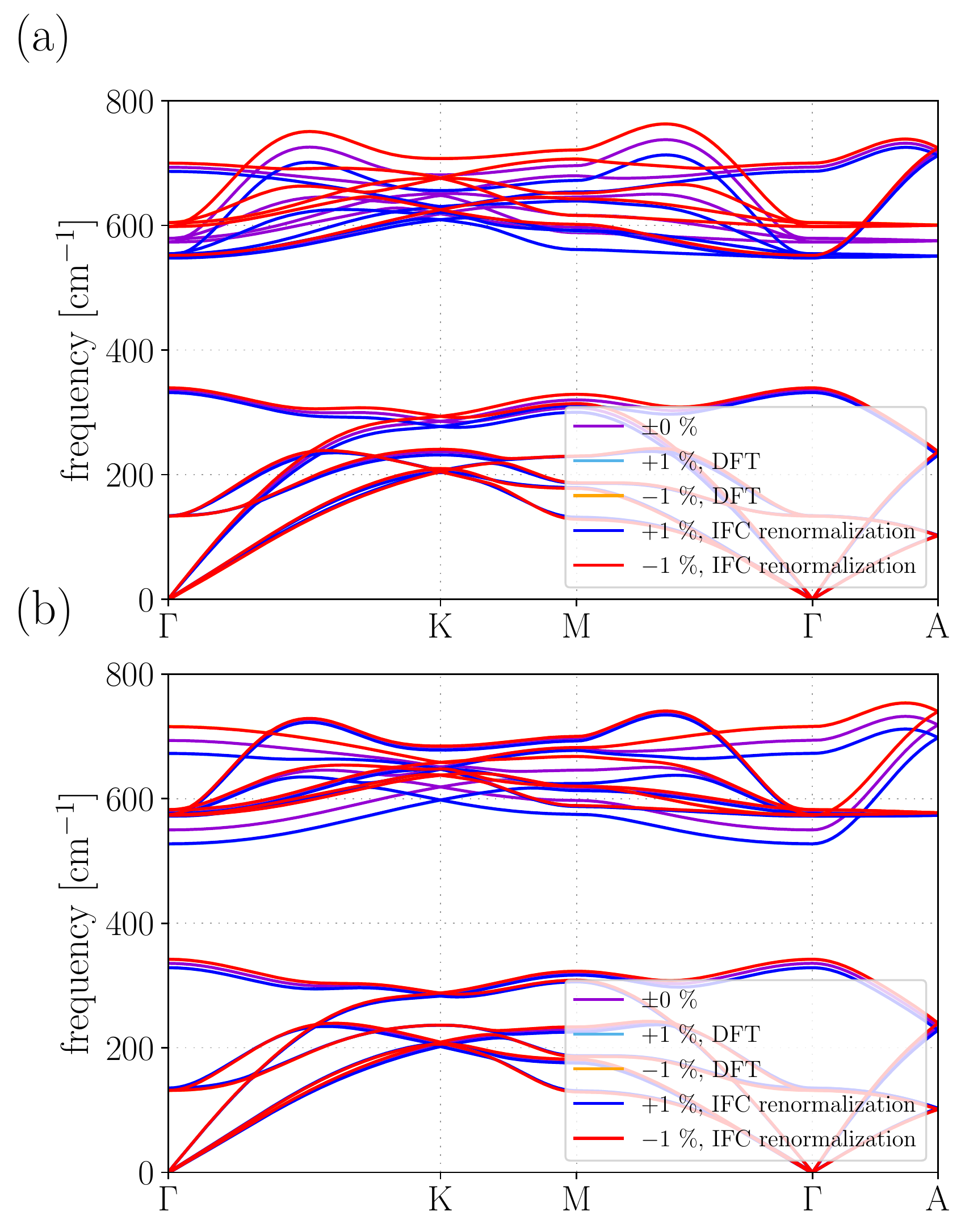}
\caption{
The change of the harmonic phonon dispersion of GaN induced by external strains.
We compare the harmonic phonon dispersions calculated by the conventional frozen phonon method with DFT calculations on the deformed cells (DFT) and those obtained by the \gls{ifc} renormalization (\gls{ifc} renormalization).
(a) The case that the lattice constant $a$ is expanded by $\pm 1$ \%.
(b) The case that the lattice constant $c$ is expanded by $\pm 1$ \%.
Note that the calculation results of the \gls{ifc} renormalization overlap with those of DFT.
}
\label{Fig_GaN_IFCrenormalize_harmonic_disp}
\end{center}
\end{figure}

\begin{figure}[h]
\vspace{0cm}
\begin{center}
\includegraphics[width=0.48\textwidth]{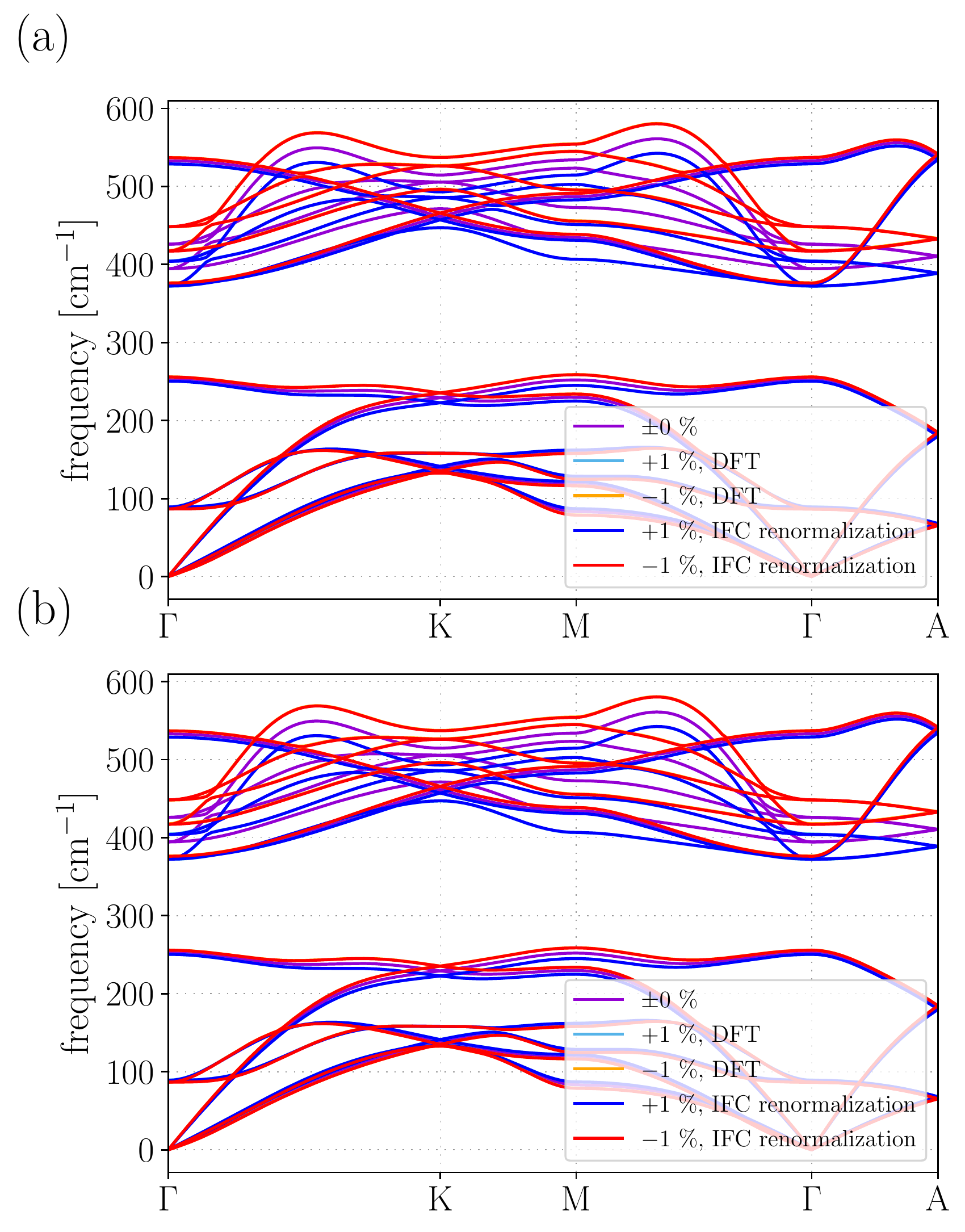}
\caption{
The change of the harmonic phonon dispersion of ZnO induced by external strains.
We compare the harmonic phonon dispersions calculated by the conventional frozen phonon method with DFT calculations on the deformed cells (DFT) and those obtained by \gls{ifc} renormalization (\gls{ifc} renormalization).
(a) The case that the lattice constant $a$ is expanded by $\pm 1$ \%.
(b) The case that the lattice constant $c$ is expanded by $\pm 1$ \%.
Note that the calculation results of the \gls{ifc} renormalization overlap with those of DFT.
}
\label{Fig_ZnO_IFCrenormalize_harmonic_disp}
\end{center}
\end{figure}

In this Appendix, we verify that the \gls{ifc} renormalization accurately reproduces the results of corresponding \gls{dft} calculations. 
We first investigate the phonon frequency shift induced by the structural change.
This is important because the thermal expansion coefficient can be rewritten using the derivatives of the phonon frequencies, which is well known as the Gr\"uneisen formula~\cite{https://doi.org/10.1002/andp.19123441202, doi:10.1063/1.5125779}.
We calculate the phonon dispersion curves of GaN and ZnO with slightly changed lattice constants using the \gls{ifc} renormalization and the conventional \gls{dft}-based frozen phonon method on the deformed unit cells.
As shown in Figs.~\ref{Fig_GaN_IFCrenormalize_harmonic_disp} and~\ref{Fig_ZnO_IFCrenormalize_harmonic_disp}, the calculation results of the two methods are almost identical, which validates the use of \gls{ifc} renormalization for calculating the thermal expansion.

\begin{figure}[h]
\vspace{0cm}
\begin{center}
\includegraphics[width=0.48\textwidth]{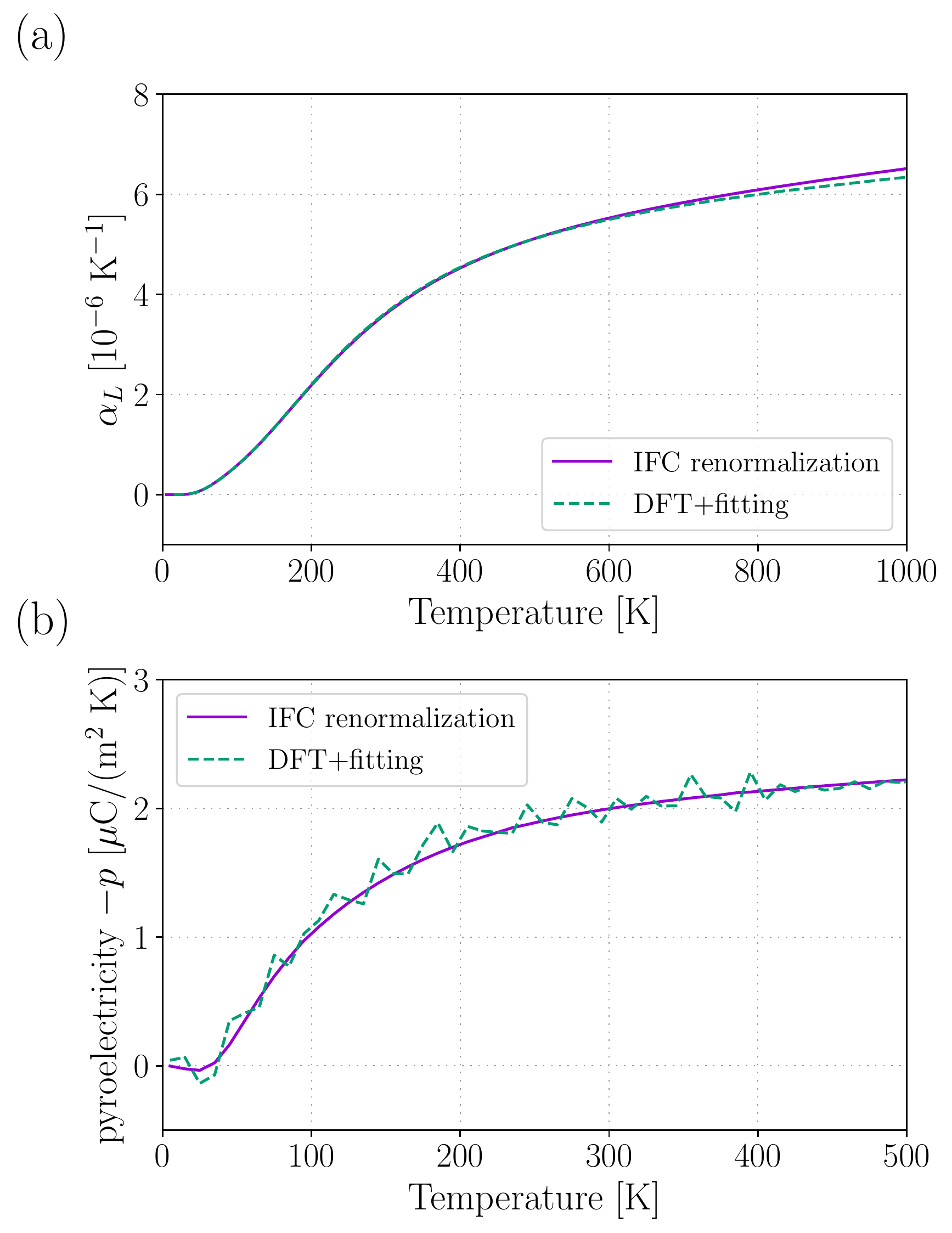}
\caption{
The comparison of \gls{qha} calculation results on wurtzite GaN that are based on \gls{ifc} renormalization (\gls{ifc} renormalization) or DFT calculations (DFT+fitting). In the DFT+fitting method, we calculate the harmonic phonon dispersion curves from \gls{dft} results for several structures and fit the free energy. We consider the single-parameter optimization problems because DFT-based \gls{qha} is computationally costly to apply to multi-parameter cases.
(a) Isotropic thermal expansion is assumed and $a(T)/a(T=0) = c(T)/c(T=0)$ is optimized while the internal coordinates are fixed. 
The $V$ dependence of the free energy is fitted by the Birch-Murnaghan equation of state~\cite{PhysRev.71.809, doi:10.1073/pnas.30.9.244} in the DFT+fitting approach.
The linear thermal expansion coefficient $\alpha_L = \frac{1}{a} \frac{da}{dT}$ is plotted. 
(b) The internal coordinates are optimized while cell shape is fixed. The optimum internal coordinates are determined by fitting the free energy by a fourth order polynomial in terms of $u^{(0)}_{\text{Ga},z} - u^{(0)}_{\text{N},z}$ in the DFT+fitting approach.
}
\label{Fig_GaN_compare_with_DFTQHA}
\end{center}
\end{figure}

\begin{figure}[h]
\vspace{0cm}
\begin{center}
\includegraphics[width=0.48\textwidth]{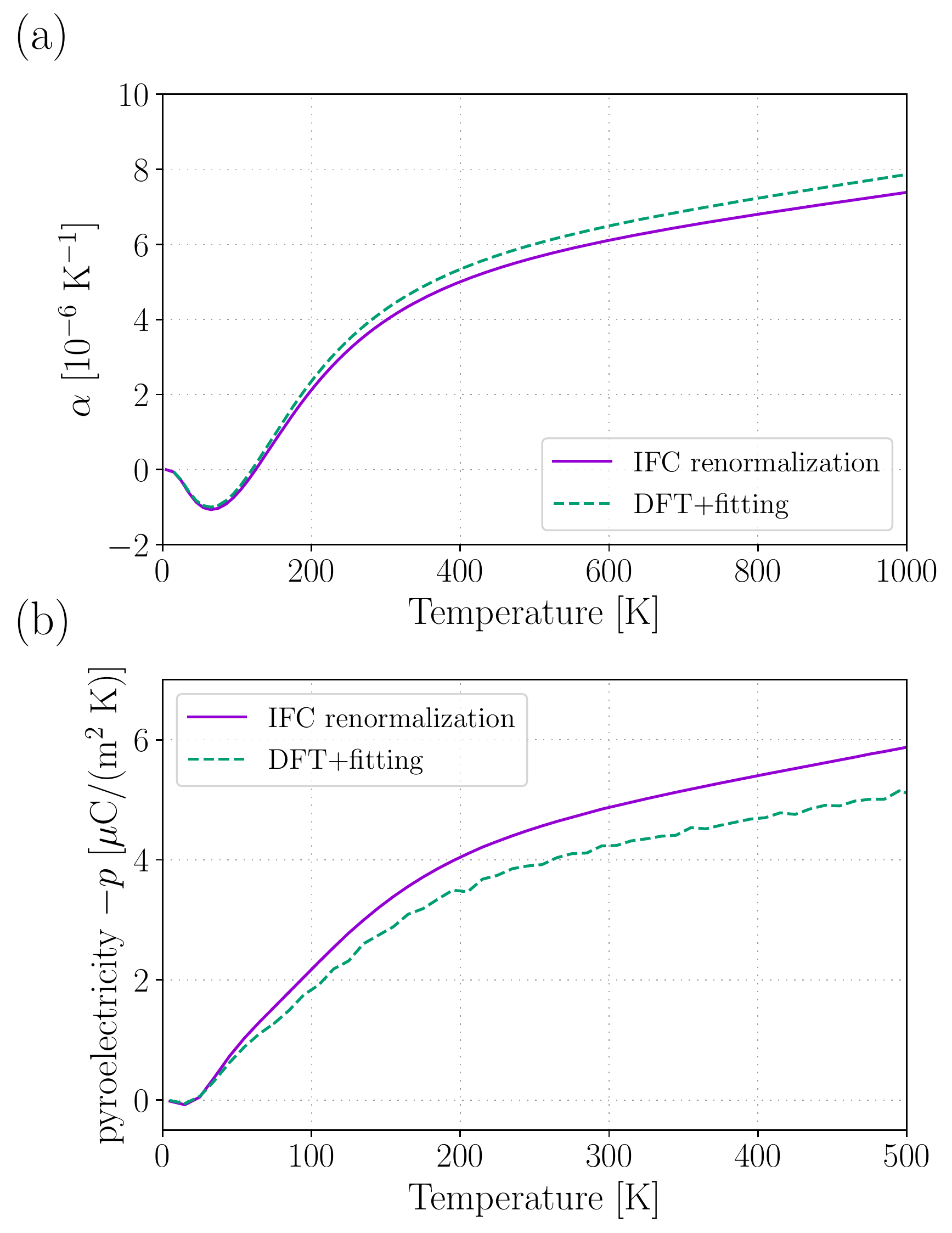}
\caption{
The comparison of \gls{qha} calculation results on wurtzite ZnO that are based on \gls{ifc} renormalization (\gls{ifc} renormalization) or DFT calculations (DFT+fitting). In the DFT+fitting method, we calculate the harmonic phonon dispersion curves from DFT results for several structures and fit the free energy. We consider the single-parameter optimization problems because DFT-based \gls{qha} is computationally costly to apply to multi-parameter cases. 
(a) Isotropic thermal expansion is assumed and $a(T)/a(T=0) = c(T)/c(T=0)$ is optimized while the internal coordinates are fixed. 
The $V$ dependence of the free energy is fitted by the Birch-Murnaghan equation of state~\cite{PhysRev.71.809, doi:10.1073/pnas.30.9.244} in the DFT+fitting approach.
The linear thermal expansion coefficient $\alpha_L = \frac{1}{a} \frac{da}{dT}$ is plotted. 
(b) The internal coordinates are optimized while cell shape is fixed. The optimum internal coordinates are determined by fitting the free energy by a fourth order polynomial in terms of $u^{(0)}_{\text{Zn},z} - u^{(0)}_{\text{O},z}$ in the DFT+fitting approach.
}
\label{Fig_ZnO_compare_with_DFTQHA}
\end{center}
\end{figure}

Subsequently, we compare the \gls{qha} results obtained by fitting the \gls{qha} free energies, which are calculated at several different structures from the \gls{dft}-based frozen phonon calculations (DFT+fitting), and the results obtained by using the \gls{ifc} renormalization.
We consider single-parameter optimizations because the DFT+fitting method is inefficient to apply to multi-parameter cases.
The calculation results are shown in Figs.~\ref{Fig_GaN_compare_with_DFTQHA} and~\ref{Fig_ZnO_compare_with_DFTQHA}. We can see that the results of the \gls{ifc} renormalization and the DFT+fitting show good agreement, especially for gallium nitride. This difference can be attributed to the fact that the fitting error in the IFC calculation is lower in GaN (0.7696 \%) than in ZnO (2.1930 \%).

In Figs.~\ref{Fig_GaN_compare_with_DFTQHA} and~\ref{Fig_ZnO_compare_with_DFTQHA}, the pyroelectricity calculated by DFT+fitting shows unphysical fluctuations. This occurs presumably because we fit the free energy by a fourth-order polynomial of the atomic displacement, which is less accurate than fitting with an equation of state for the thermal expansion coefficient. As the equation of state generally considers free energy as a function of volume and temperature~\cite{PVinet_1989, PhysRev.71.809, doi:10.1073/pnas.30.9.244}, the fitting procedure can be problematic for multi-parameter optimization or optimization of internal coordinates with the DFT+fitting approach of \gls{qha}.

\bibliography{apssamp}% Produces the bibliography via BibTeX.

\end{document}